\documentclass[conference]{IEEEtran}
\IEEEoverridecommandlockouts

\usepackage{cite}
\usepackage{amsmath,amssymb,amsfonts}

\usepackage[frozencache,cachedir=minted-cache]{minted}
\usepackage{algorithmic}
\usepackage{graphicx}
\usepackage{textcomp}
\usepackage{xcolor}
\def\BibTeX{{\rm B\kern-.05em{\sc i\kern-.025em b}\kern-.08em
    T\kern-.1667em\lower.7ex\hbox{E}\kern-.125emX}}

\usepackage{booktabs}
\usepackage{array}
\usepackage{tabularx}
\usepackage{float}
\usepackage{xcolor}
\usepackage[margin=0.5in]{geometry} 
\usepackage[many]{tcolorbox}    		    
\usepackage{setspace}             
\usepackage{multicol}             
\definecolor{main}{HTML}{adbde3}   
\definecolor{sub}{HTML}{ebebeb}     
\usepackage{url}
\usepackage{multirow}
\usepackage{placeins}
\usepackage{hyperref}
\usepackage{makecell}
\usepackage{orcidlink}
\usepackage{threeparttable}
\usepackage{enumitem}
\usepackage{fontawesome}
\setminted[bash]{autogobble=true,breaklines=true,fontsize=\footnotesize}
\setminted[python]{autogobble=true,breaklines=true,fontsize=\footnotesize}
\makeatletter 
\let\MYcaption\@makecaption
\makeatother
\usepackage[font=footnotesize]{subcaption}

\makeatletter
\let\@makecaption\MYcaption
\makeatother

\usepackage{siunitx}
\DeclareSIUnit\flop{\textsc{Fl}\textsc{Op}}
\DeclareSIUnit\cycle{\textsc{Cycle}}
\DeclareSIUnit[per-mode=symbol]\floppersec{\flop\per\second}
\DeclareSIUnit[per-mode=symbol]\flopperjoule{\flop\per\joule}
\DeclareSIQualifier{\doubleprecision}{FP64}
\DeclareSIQualifier{\singleprecision}{FP32}
\DeclareSIQualifier{\halfprecision}{FP16}
\DeclareSIQualifier{\theoretical}{th}
\DeclareSIUnit\pixel{px}
\definecolor{fzjblue}{RGB}{2,61,107} 

\newcommand{\jpwr}{jpwr}

\begin{document}
\newtcolorbox{boxE}{
    colback = white,
    enhanced,
    boxrule = 0pt, 
    borderline = {0.75pt}{0pt}{main}, 
    borderline = {0.75pt}{2pt}{sub} 
}

\title{Performance and Power: Systematic Evaluation of AI Workloads on Accelerators with CARAML}

\author{\IEEEauthorblockN{1\textsuperscript{st} Chelsea Maria John}
\IEEEauthorblockA{\textit{Jülich Supercomputing Centre} \\
\textit{Forschungszentrum Jülich}\\
Jülich, Germany \\
c.john@fz-juelich.de}
\and
\IEEEauthorblockN{Stepan Nassyr}
\IEEEauthorblockA{\textit{Jülich Supercomputing Centre} \\
\textit{Forschungszentrum Jülich}\\
Jülich, Germany \\
s.nassyr@fz-juelich.de}
\and
\IEEEauthorblockN{Carolin Penke}
\IEEEauthorblockA{\textit{Jülich Supercomputing Centre} \\
\textit{Forschungszentrum Jülich}\\
Jülich, Germany \\
c.penke@fz-juelich.de}
\and
\IEEEauthorblockN{Andreas Herten}
\IEEEauthorblockA{\textit{Jülich Supercomputing Centre} \\
\textit{Forschungszentrum Jülich}\\
Jülich, Germany \\
a.herten@fz-juelich.de}
}
\author{
    \IEEEauthorblockN{
        Chelsea Maria John\,\orcidlink{0000-0003-3777-7393},
        Stepan Nassyr\,\orcidlink{0000-0002-0035-244X},
        Carolin Penke\,\orcidlink{0000-0002-4043-3885},
        Andreas Herten\,\orcidlink{0000-0002-7150-2505}
    }
    \IEEEauthorblockA{
        \textit{J\"{u}lich Supercomputing Centre} \\
        \textit{Forschungszentrum J\"{u}lich} \\
        J\"{u}lich, Germany
    }
}

\maketitle
\begin{abstract}
The rapid advancement of machine learning (ML) technologies has driven the development of specialized hardware accelerators designed to facilitate more efficient model training. This paper introduces the CARAML benchmark suite, which is employed to assess performance and energy consumption during the training of transformer-based large language models and computer vision models on a range of hardware accelerators, including systems from NVIDIA, AMD, and Graphcore. CARAML provides a compact, automated, extensible, and reproducible framework for assessing the performance and energy of ML workloads across various novel hardware architectures. The design and implementation of CARAML, along with a custom power measurement tool called \jpwr, are discussed in detail.
\end{abstract}

\begin{IEEEkeywords}
Machine Learning, Energy, NLP, Computer Vision, AI, Performance Measurement, Benchmark, GPU, IPU, Accelerators
\end{IEEEkeywords}

\section{Introduction}
Fueled by the growing interest in training ever larger deep neural networks, such as large language models and other foundation models, the demands for hardware specialized on these workloads have grown immensely. Graphics processing units (GPUs) have evolved from their origins in computer graphics to become the primary computational engines of the AI revolution. While the central processing unit (CPU) controls a program's execution flow, it offloads compute-intensive highly-parallel tasks to the GPU (the \emph{accelerator}). Evolving from a pioneering company, NVIDIA has emerged as the dominant player in the market as of 2024, spearheading current hardware developments. Other vendors, such as AMD and Intel, also provide GPUs aiming to accelerate model training and inference. 

Another promising class of AI accelerators is based on the idea of distributed local per-compute-unit memory together with on-chip message passing, in contrast to a shared memory hierarchy, typical to classical CPUs and GPUs. Vendors following this \emph{dataflow} approach include Graphcore, Cerebras, Groq, SambaNova, and Tenstorrent. 

Performance characteristics not only vary between generations and vendors, but depend on the node or cluster configuration in which the accelerator is embedded, including CPU, memory, and interconnect. When evaluating and comparing these heterogeneous hardware options, e.g. for purchase decisions in an academic or industrial setting, it is not  sufficient to compare hardware characteristics such as number of cores, thermal design power (TDP), theoretic bandwidth, or peak performance in \qty[per-mode=symbol]{}{\floppersec}. Their effect on workload performance is not straightforward, and the accelerator architectures might barely be comparable. Performance data reflecting the actual intended workloads, collected on various competing systems independently of vendor interests, offer highly valuable information. Power consumption is one such important metric in this regard.

In particular within the field of machine learning, having a structured, automatic benchmarking tool to investigate the effect of hyperparameters, such as learning rate and batch size, and to identify optimal settings is important. When training large models across multiple cluster nodes, additional hyperparameters are necessary to define the parallelization layout, leveraging various forms of parallelism. Moreover, hardware configurations, such as those related to processor affinity or network communication, must be systematically explored.

As a framework to collect this data, we present the CARAML benchmark suite, a \textbf{C}ompact, \textbf{A}utomated, \textbf{R}eproducible \textbf{A}ssessment of \textbf{M}achine-\textbf{L}earning workloads on novel accelerators. Further, we contribute performance data on various accelerators, collected with CARAML, including energy measurements, as well as first-hand experiences with encountered challenges. The machine learning workloads include the training of a Generative Pretrained Transformer~(GPT)-based large language model using \texttt{PyTorch}, as well as training of a \texttt{ResNet50} model implemented in \texttt{TensorFlow}. Energy measurements are facilitated by our \jpwr{} tool, which we also contribute with this paper.

The remaining paper is structured as follows: in \autoref{Sec:Background} we provide background information on the state-of-the-art AI workloads included in CARAML in the fields of natural language processing and image recognition. Details on the investigated hardware configurations. More details on the implementation of the benchmark, including a description of used tools, requirements, and execution instructions, are presented in \autoref{Sec:Benchmark}. Performance results, in terms of throughput (\qty[per-mode=symbol]{}{images\per\second}, \qty[per-mode=symbol]{}{tokens\per\second}) for various batch sizes, are reported in \autoref{Sec:Results}. Additional energy metrics such as \qty[per-mode=symbol]{}{images\per Wh} and \qty[per-mode=symbol]{}{tokens\per Wh} provide another layer of insight. In \autoref{Sec:TechnicalDifficulties}, we briefly discuss the difficulties encountered in the benchmark development in terms of hardware and software compatibility and comparability, and how they are resolved. Final conclusions, take-away messages, and future plans are given in \autoref{Sec:Conclusions}.

\section{Background}\label{Sec:Background}

Deep learning constitutes an important workload of high performance computing (HPC) clusters, only increasing in significance in recent years. A neural network aims to generalize an input-output relation observed in its training data. A chain of matrix products and activation functions, reflecting a neural network architecture, is applied to the batched input data in a forward pass. The backward pass, also called back-propagation, updates the matrix elements (the “weights”) using further matrix products, This operation reflects a step of (stochastic) gradient descent to minimize the loss function representing the difference between computed and true output of the sample. 

Although specialized network architectures like transformers and convolutional neural networks have been developed for specific tasks, they rely on the fundamental building block of matrix multiplication. Since matrix multiplications are inherently parallel, many-core hardware architectures, such as GPUs and other accelerators optimized for this process, are crucial.
\subsection{Natural Language Processing }\label{Sec:Background:NLP}
The advent of large language models based on transformer architectures \cite{vaswani2017attention} has revolutionized the field of natural language processing (NLP). These models have enabled significant advancements across a wide range of tasks, including speech recognition, text classification, natural language understanding, and generation. By leveraging shared foundational models, these tasks can now be effectively addressed with additional fine-tuning or in-context learning, allowing for greater flexibility and efficiency in various domains.

Language models use large text corpora as training data to predict the next piece of text (a \emph{token}) given the preceding context. The original transformer architecture consists of an encoder and a decoder connected via a cross-attention mechanism. Attention is a key operation in this architecture, characterized by its quadratic complexity in the sequence length. It involves matrix-matrix products of learned token representations, allowing the model to capture relationships between tokens while accounting for their relative positions.

Scaling up transformer models is achieved by stacking multiple transformer layers, each built around an attention mechanism, along with feed-forward layers, residual connections, and normalization. This increases the number of learnable parameters, making larger networks more capable, especially when trained on sufficiently large datasets. To handle the computational demands of such large models, parallelization techniques are essential. State-of-the-art methods include 3D parallelism \cite{narayanan2021efficientlargescalelanguagemodel, john2023opengpt} (combining data, tensor, and pipeline parallelism), sequence parallelism \cite{korthikanti2023reducing}, activation recomputation \cite{korthikanti2023reducing}, and optimizations like flash attention \cite{dao2024flashattention}. 

The NLP benchmark task in CARAML utilizes the Megatron-LM framework \cite{shoeybi2020megatronlmtrainingmultibillionparameter,narayanan2021efficientlargescalelanguagemodel}, a robust, research-focused software platform developed by NVIDIA using \texttt{PyTorch}. Megatron-LM has been instrumental in advancing large-scale language model training, incorporating and pioneering the previously mentioned features. The BigCode Project fork of Megatron-LM with ROCm adaptations is used for AMD and forked version of Graphcore application examples is used on Graphcore.

\newcommand{\iconPerf}{\includegraphics[height=2ex]{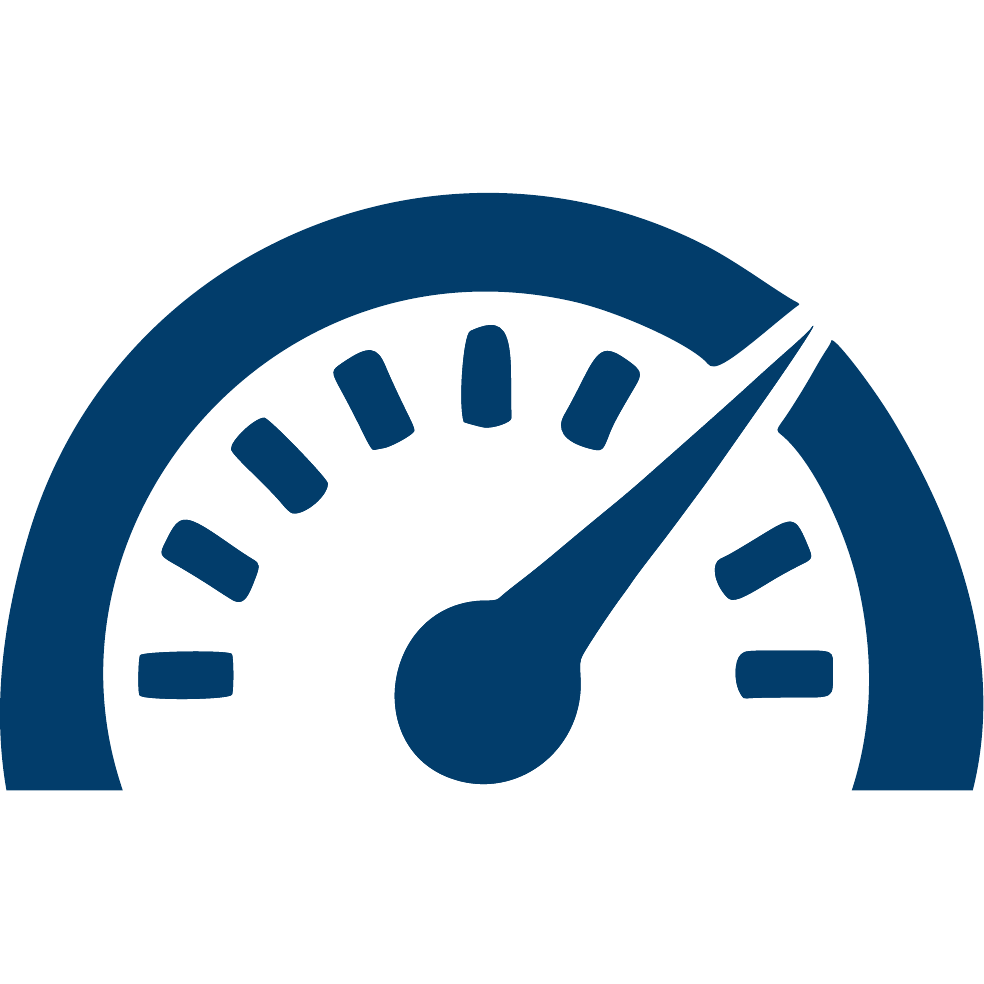}}
\newcommand{\iconSM}{\includegraphics[height=2ex]{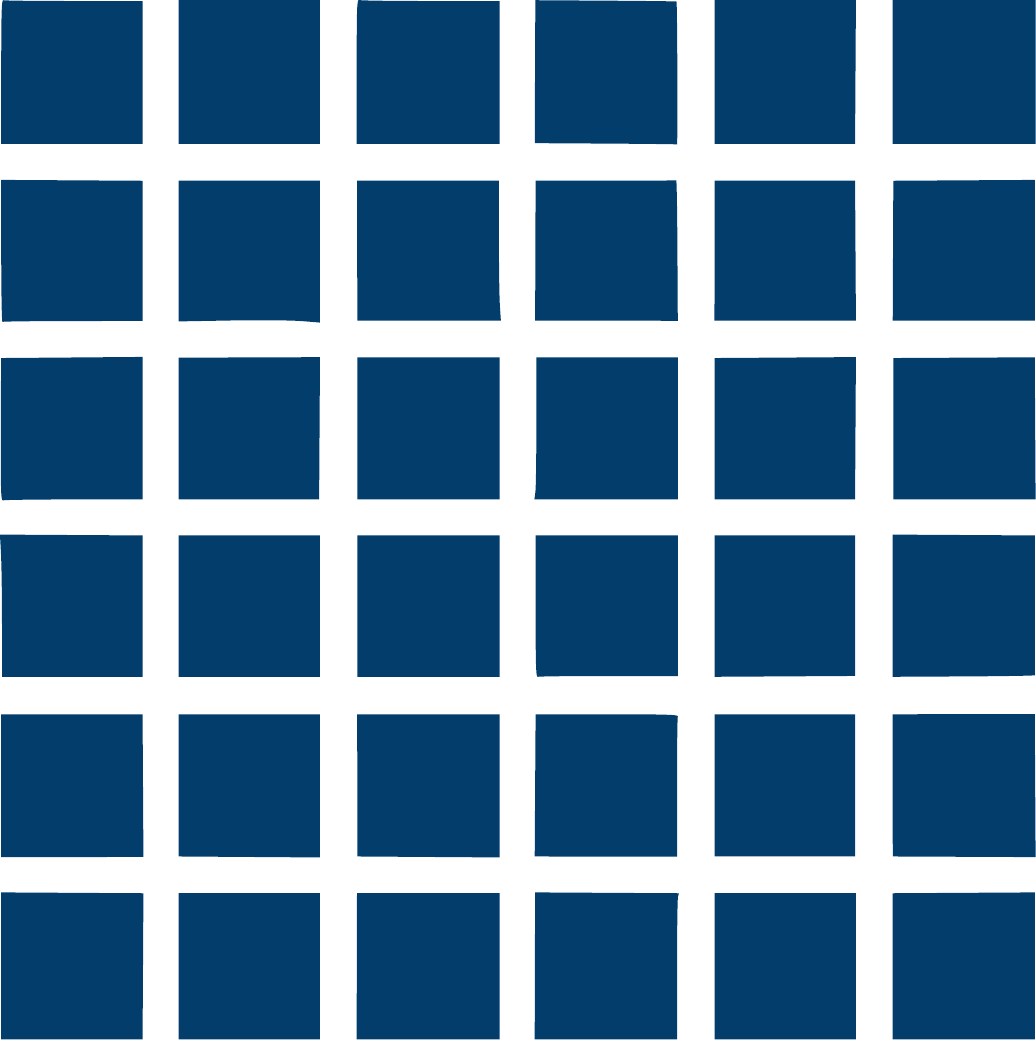}}
\newcommand{\iconMem}{\includegraphics[height=2ex]{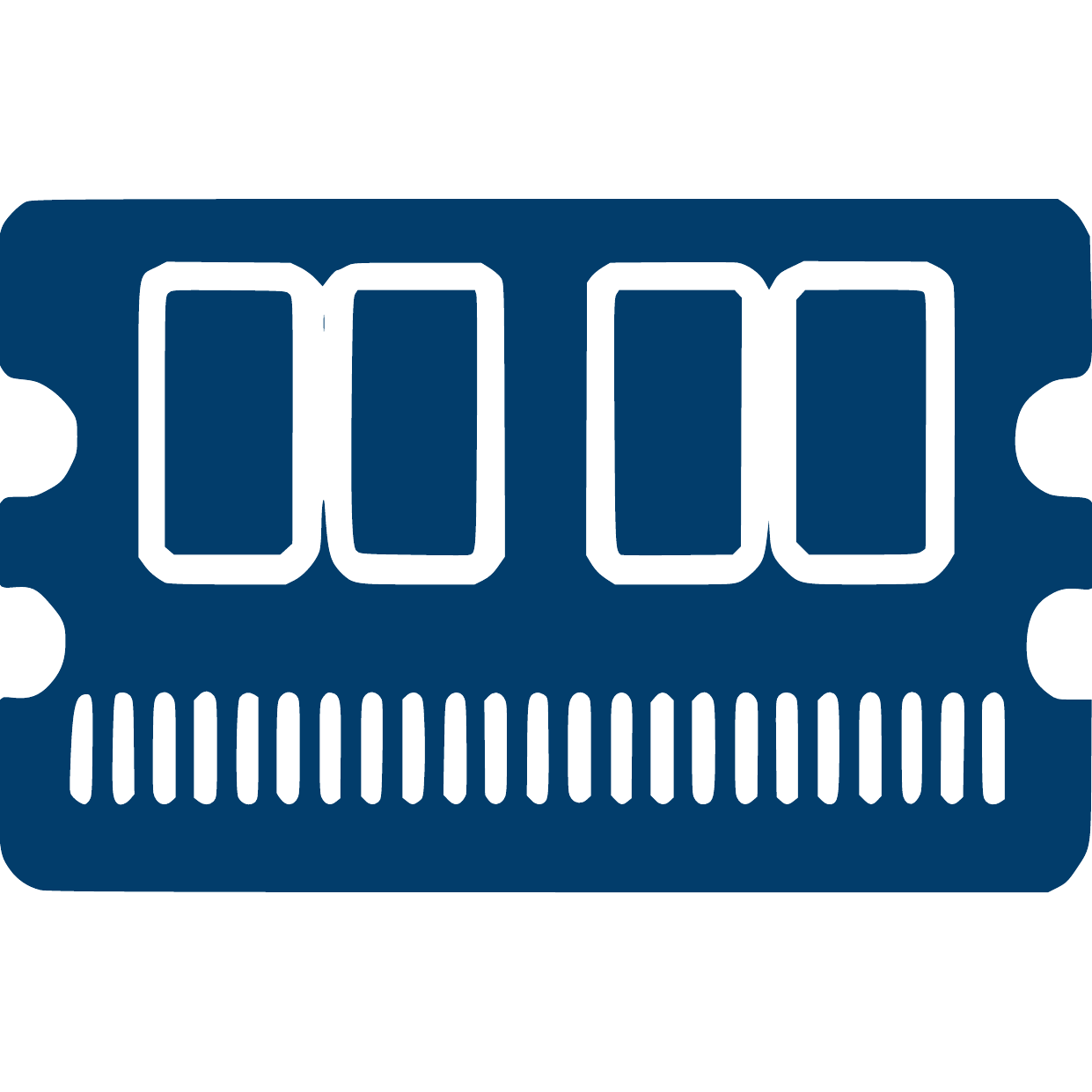}}
\newcommand{\iconCPU}{\includegraphics[height=2ex]{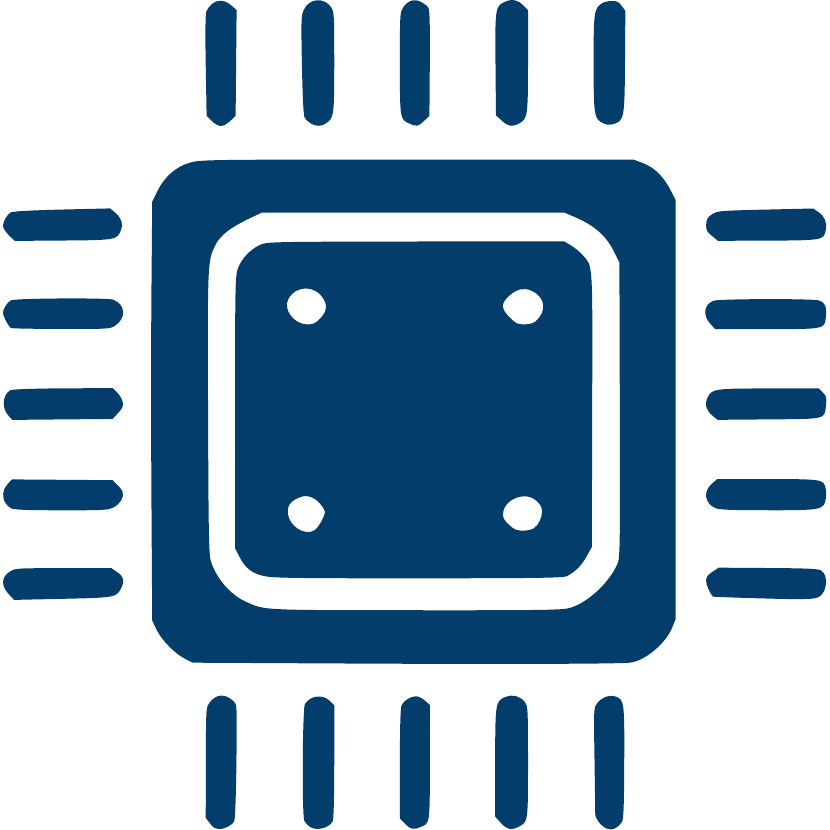}}
\begin{figure}[t]
\small
\begin{boxE}
\begin{itemize}[leftmargin=*]
    \item \textbf{NVIDIA A100 GPU (SXM4)}:
        \begin{itemize}[leftmargin=*]
            \item[\iconSM] 108 SM (each: 64 CUDA cores, 4 Tensor Cores)
            \item[\iconPerf] Peak performance FP16: \qty[per-mode=symbol]{312}{\tera\floppersec}
            \item[\iconMem] Memory: \qty{40}{\giga\byte} HBM2e
        \end{itemize}
    \item \textbf{NVIDIA H100 GPU (PCIe)
    }: 
    \begin{itemize}[leftmargin=*]
        \item[\iconSM] 114 SM (each: 128 CUDA cores, 4 Tensor Cores)
        \item[\iconPerf] Peak performance FP16: \qty[per-mode=symbol]{756}{\tera\floppersec}  
        \item[\iconMem] Memory: \qty{80}{\giga\byte} HBM2e
    \end{itemize}
    \item \textbf{NVIDIA H100 GPU (SXM5)
    }
        \begin{itemize}[leftmargin=*]
        \item[\iconSM] 132 SM (each: 128 CUDA cores, 4 Tensor Cores)
        \item[\iconPerf] Peak performance FP16: \qty[per-mode=symbol]{990}{\tera\floppersec}
        \item[\iconMem] Memory: \qty{94}{\giga\byte} HBM2e
        \end{itemize}
    \item \textbf{NVIDIA GH200 Superchip}: 
    \begin{itemize}[leftmargin=*]
        \item[\iconCPU] CPU: NVIDIA Grace (Arm Neoverse-V2), 72 cores 
        \item[\iconSM] GPU: NVIDIA Hopper H100,	132 SM (each: 128 CUDA cores, 4 Tensor Cores)
        \item[\iconPerf] Peak performance FP16 : \qty[per-mode=symbol]{990}{\tera\floppersec}
        \item[\iconMem] Memory: \qty{96}{\giga\byte} HBM3 (GPU) at \qty{4}{\tera\byte\per\second}, up to \qty{480}{\giga\byte} LPDDR5X (CPU) at up to \qty{512}{\giga\byte\per\second} (\qty{8532}{\mega\hertz})
    \end{itemize}
    \item \textbf{AMD MI250 GPU}:
   \begin{itemize}[leftmargin=*]
        \item[\color{fzjblue}\faExclamationCircle] GPU built from two Graphics Compute Dies (GCD) as Multi-Chip Module (MCM). The numbers below refer to one MCM, i.e. two logical GPUs as seen by the operating system.
        \item[\iconSM] $2\times$ 104 CU (each: 64 stream processors, 4 Matrix Cores)
        \item[\iconPerf] Peak performance FP16: \qty[per-mode=symbol]{362.1}{\tera\floppersec}
        \item[\iconMem] Memory: \qty{128}{\giga\byte} HBM2e
   \end{itemize} 
   \item \textbf{Graphcore GC200 IPU}
   :  
    \begin{itemize}[leftmargin=*]
        \item[\iconSM] \num{1472} processor cores (\emph{IPU cores})
        \item[\iconPerf] Peak performance FP16: \qty[per-mode=symbol]{250}{\tera\floppersec}
        \item[\iconMem] Memory: \qty{900}{\mega\byte}, distributed to cores
    \end{itemize}
\end{itemize}
\end{boxE}
\caption{List of evaluated accelerators. NVIDIA's \emph{Streaming Multiprocessors} are abbreviated with \emph{SM}. AMD's \emph{Compute Units} are abbreviated with \emph{CU}. Peak performance is given without sparsity.}
\label{Fig:Accelerators}
\end{figure}

\subsection{Computer Vision}
Computer vision (CV), particularly its core task of image classification, has been a significant driver of deep learning advancements due to its numerous real-world applications. The architecture of convolutional neural networks (CNNs) effectively addresses challenges such as vanishing and exploding gradients by employing successive layers of convolutional filters, defined by learned parameters. It became evident that the massively parallel architecture of GPUs is well-suited for the computational demands of CNNs, leading to substantial improvements in training efficiency and model performance \cite{chellapilla:inria-00112631}.

Residual connections were introduced by the ResNet model family \cite{heDeepResidualLearning} and solved the problem of \emph{degradation}, i.e. the counterintuitive observation of higher training errors in deeper networks with more parameters. 

While the transformer architecture has also found its way into image recognition \cite{dosovitskiy2021imageworth16x16words}, convolutional neural networks are extremely mature and widely used in production environments, and represent an important benchmark case. Together with transformers, they cover a wide portion of currently relevant deep learning paradigms.

In CARAML, a benchmark to train a  \texttt{ResNet50} model, i.e. a ResNet model with 50 convolutional layers from scratch, is curated from a forked version of the official TensorFlow CNN benchmark. The benchmark employs data-parallelism with \texttt{Horovod} to use multiple GPUs. In data-parallelism, each device holds a model copy but performs a backpropagation for a different batch of input data, combining the gradients after each step using an all-reduce collective operation. 

\subsection{Accelerators}

\begin{table*}[t]
\begin{threeparttable}
    \centering
    \caption{Systems analyzed with CARAML. CPU cores are given as \emph{72 c} for 72 cores.}
    \footnotesize
    \renewcommand{\arraystretch}{1.5} 
    \begin{tabularx}{\linewidth}{p{2cm}X X X X X X  X}
        \toprule
        \bfseries Platform 
            & \makecell{\bfseries GH200\\JEDI} 
            & \makecell{\bfseries GH200\\JURECA} 
            & \makecell{\bfseries H100\\JURECA}
            & \makecell{\bfseries H100\\WestAI}
            & \makecell{\bfseries MI200\\JURECA} 
            & \makecell{\bfseries IPU-M2000\\JURECA} 
            & \makecell{\bfseries A100\\JURECA} \\
        \midrule
        \textbf{Accelerator} 
            & \multirow{2}{*}[-2em]{\shortstack[c]{$4\times$ NVIDIA\\GH200-120GB\\($1\times$ 72 c Grace,\\ $1\times$ H100)}} 
            & \multirow{2}{*}[-2em]{\shortstack[c]{$1\times$ NVIDIA\\GH200-480GB\\($1\times$ 72 c Grace,\\ $1\times$ H100)}}  
            & $4\times$ NVIDIA H100 GPU (PCIe) 
            & $4\times$ NVIDIA H100 GPU (SXM5)
            & $4\times$ AMD MI250 GPU (OAM)
            & $4\times$ Graphcore GC200 IPU 
            & $4\times$ NVIDIA A100 GPU (SXM4)\\
             \cmidrule{1-1}\cmidrule{4-8}
        \textbf{CPU} &   &  
            & $2\times$ 72 c Intel Xeon Platinum 8452Y
            & $2\times$ 32 c Intel Xeon Platinum 8462Y
            & $2\times$ 48 c AMD EPYC 7443
            & $2\times$ 48 c AMD EPYC 7413
            & $2\times$ 64 c AMD EPYC 7742\\
         \midrule
        \textbf{CPU--Acc. Connect (intra-node)}
            & \multicolumn{2}{c}{NVLink-C2C \qty{900}{\giga\byte\per\second}} 
            & \multicolumn{2}{c}{PCIe Gen 5 \qty{128}{\giga\byte\per\second}} 
            & \multicolumn{3}{c}{PCIe Gen 4 \qty{64}{\giga\byte\per\second}}
            \\
         \midrule
        \textbf{Acc.--Acc. Connect (intra-node)}\tnote{1} 
            & NVLink4 \qty{900}{\giga\byte\per\second} 
            & - 
            & NVLink4\tnote{2} \qty{600}{\giga\byte\per\second}
            & NVLink4 \qty{900}{\giga\byte\per\second}
            & Infinity Fabric \qty{500}{\giga\byte\per\second} 
            & IPU-Link\tnote{3} \qty{256}{\giga\byte\per\second}
            & NVLink3 \qty{600}{\giga\byte\per\second} \\
         \midrule
        \textbf{Interconnect internode}\tnote{4}  
            & $4\times$ IB NDR ($4\times$\qty{200}{\giga\bit\per\second})
            & -
            & -
            & $2\times$ IB NDR ($2\times$\qty{400}{\giga\bit\per\second})
            & $2\times$ IB HDR ($2\times$\qty{200}{\giga\bit\per\second}) 
            & -
            & $2\times$ IB HDR ($2\times$\qty{200}{\giga\bit\per\second})\\
        \midrule
        \textbf{Memory} 
            & $4\times$  \qty{120}{\giga\byte} LPDDR5X (CPU), $4\times$  \qty{96}{\giga\byte} HBM3 (GPU) 
            & \qty{480}{\giga\byte} LPDDR5X (CPU), \qty{96}{\giga\byte} HBM3 (GPU)
            & \qty{512}{\giga\byte} DDR5-4800 (CPU), \qty{80}{\giga\byte} HBM2e (GPU) 
            & \qty{512}{\giga\byte} DDR5-4800 (CPU), \qty{94}{\giga\byte} HBM2e (GPU)
            & \qty{512}{\giga\byte} DDR4-3200 (CPU), \qty{128}{\giga\byte} HBM2e (GPU)
            & \qty{512}{\giga\byte} DDR4-3200 (CPU) 
            & \qty{512}{\giga\byte} DDR4-3200 (CPU), $4\times$ \qty{40}{\giga\byte} HBM2e (GPU) \\
        \midrule
        \textbf{TDP / device}
            & \qty{680}{\watt}\tnote{\dag}
            & \qty{700}{\watt}\tnote{\dag}
            & \qty{350}{\watt}
            & \qty{700}{\watt}
            & \qty{560}{\watt}
            & \qty{300}{\watt}
            & \qty{400}{\watt} \\
        \midrule
        \textbf{JUBE Tag} 
            & \texttt{JEDI} 
            & \texttt{GH200} 
            & \texttt{H100} 
            & \texttt{WAIH100} 
            & \texttt{MI250} 
            & \texttt{GC200} 
            & \texttt{A100}\\
        \bottomrule
    \end{tabularx}
    \begin{tablenotes}
        \item[1] Bidirectional bandwidths per device.
        \item[2] GPU0 and GPU1 and GPU2 and GPU3 are connected through NVLink bridges, each with 12 NVLink4 connections (each \qty{25}{\giga\byte\per\second}).
        \item[3] Each IPU in a node is connected to other IPUs in- and out-of-node with \num{10} \emph{IPU-Links}. Intra-node, an IPU connects to two other IPUs with \num{2} links, and with one IPU with \num{4} links. At \qty{32}{\giga\byte\per\second} bidirectional bandwidth per link, an IPU has hence an accumulated intra-node connection bandwidth of \qty{256}{\giga\byte\per\second}.
        \item[4] NVIDIA InfiniBand is abbreviated to \emph{IB}.
        \item[\dag] The TDP for the GH200 superchips is for the full package, i.e. including the CPU and GPU devices.
    \end{tablenotes}
    \label{Tab:Systems}
\end{threeparttable}
\end{table*}

GPUs have become the standard hardware for accelerating neural network training. However, while processing power has advanced rapidly in recent years, memory bandwidth has not kept pace, leading to potential bottlenecks. Accelerators based on data-flow architectures, such as Graphcore IPUs \cite{jiaDissectingGraphcoreIPU2019}, offer a promising alternative by addressing these limitations. Unlike traditional architectures that rely on a shared memory hierarchy, IPUs leverage distributed per-core memory, which allows for faster data loading due to the physical proximity of memory to each core. This design enables all cores to operate independently, making the processing of irregular or sparse neural network architectures more efficient. In terms of Flynn's taxonomy \cite{FlynnTaxonomy}, this can be considered a MIMD (multiple instruction streams, multiple data streams) architecture, while GPUs follow a SIMD (single instruction stream, multiple data streams) approach. 

Figure \ref{Fig:Accelerators} provides a list of accelerators that have been explored in this work utilizing the CARAML benchmarks. The complete node configurations, including CPU, memory and interconnect, are documented in Table \ref{Tab:Systems}.

We examine two generations of NVIDIA GPUs (A100 and H100) in various configurations. The A100 node is part of the JURECA DC~\cite{Thornig2021-yw} cluster at Jülich Supercomputing Centre, while the other nodes are part of the JURECA evaluation platform~\cite{jureca-eval} or the WestAI~\cite{WestAI} cluster. The two H100 node variants mainly differ in the CPU model, intra-node bandwidth and the amount of GPU memory.  

The NVIDIA GH200 superchip is built up from a Grace CPU and a Hopper GPU that are connected on chip by a high-bandwidth, low-latency interconnect. The GPU can directly access the CPU memory without explicit transfers, potentially accelerating hybrid workloads. The two GH200 nodes we investigate differ in their configuration: A node of the JEDI system contains 4 NVIDIA GH200s, while the GH200 node of the JURECA evaluation platform only has one GH200.The amount of per-node memory is the same in both cases.

The AMD MI200 node from the JURECA evaluation platform contains four MI250 GPUs. Similar to the GH200, it can be seen as combining multiple devices on a single chip. Each MI250 contains two Graphics Compute Dies, that are seen as a GPU by the operating system. From that viewpoint, each node would contain 8 GPUs. Each pair of devices may have a different transfer bandwidth \cite{Herten:916416}.

There are many programming paradigms with different compatibility characteristics \cite{herten23ManyCoresProgramminModels}. \texttt{CUDA} is well-established and serves as a blueprint for other programming models (HIP), or as a backend for portability layers such as  \texttt{Open\-ACC} or \texttt{SYCL}.  While the software ecosystems for data-flow architectures are growing, they are not yet as mature and supported by third-party software as GPU programming models. A clear leader or common standard among the competitors has not yet been established. Graphcore systems can be instructed using the \texttt{Poplar} software development kit. At a higher level, machine learning frameworks such as \texttt{PyTorch} and \texttt{TensorFlow} can be seen as portability layers, serving various platforms via vendor-specific backends. This approach of using a common codebase for various architectures is followed by the benchmarks in this paper wherever possible. The limits of this approach are described in section~\ref{Sec:TechnicalDifficulties}.

\subsection{Related Work}
Since the advent of modern computing technologies, benchmarks have been an important tool to assess the performance of hardware systems, spanning from consumer devices\cite{userbenchmark} and accelerators\cite{Juckeland15SPEC},  to  HPC clusters as major research facilities\cite{herten24JuBench}. Synthetic benchmarks, which concentrate on specific yet commonly used compute patterns, are valuable in this context \cite{asanovic09dwarfs}. However, the performance metrics they provide can be difficult to apply to more complex, real-world applications. 

Benchmarks that incorporate the workloads of real applications, or their variations, are thus extremely valuable for evaluating hardware capabilities. They also offer a way to assess the effects of parameter choices or code optimizations.

In the field of machine learning, the MLPerf set of benchmark suites \cite{MLPerfSuite} is an established industry standard supported by all major vendors. Various MLPerf suites focus on training at device or cluster level\cite{MLSYS2020_411e39b1,Farrell21MLPerf}, or on performing inference across various devices \cite{MLperfInference}. Results are collected as a coordinated effort in a yearly industry-wide competition. Based on the premise of lacking (performance) portability, vendors are expected to port and optimize a reference code for their architecture, showcasing its capabilities. Established competition rules act as clearly defined guardrails and make a comparison possible. In this competitive context, the choice of time-to-solution as a benchmark metric over throughput-based metrics makes sense, as the latter ones could be optimized at the expense of the first one. The downside of the time-to-solution metric, which here refers to the runtime until a specified accuracy is achieved, is its high computational cost. 

Similar to MLPerf, the SPEC benchmarks \cite{Juckeland15SPEC} are a consortium-driven effort to benchmark the performance of hardware systems in terms of general-purpose algorithms. It's closed source code is not freely available and does not contain ML specific workloads. 

CARAML on the other hand focuses on the user rather than the vendor perspective. As a free and open source framework under a permissive MIT license, it empowers users to evaluate the out-of-the-box performance of accelerators with minimal code adaptions. It relies on two widely used machine learning frameworks (\texttt{PyTorch} and \texttt{TensorFlow}) as portability layers. Focusing on throughput and performance in images or tokens per second allows for quick evaluation without the need to perform full training runs, even with limited computational resources. This resource-efficiency and immediate feedback, together with CARAML's high level of automation, allows its user to rapidly explore an architecture's (hyper-) parameter space or to perform parameter ablation studies.

Recent years have seen growing efforts to minimize environmental impact of HPC systems, motivated by an ongoing climate crisis and a changing energy landscape. To this end, measuring energy consumption of HPC hardware and workloads has come more into focus. Efforts to assess energy efficiency on a cluster level \cite{Feng2007TheGL} are now accompanied by the development of tools for a more fine-grained assessment \cite{Muriedas2023Perun}. AI workloads such as large language models in particular have come under scrutiny due to the significant energy footprint required for their training \cite{patterson2021carbonemissionslargeneural, Luccioni24Carbon}. 
The \jpwr{} tool was developed as a compact prototype to fulfil the requirements of incorporating energy measurement in the CARAML benchmark suite.

\section{The CARAML benchmark}\label{Sec:Benchmark}

The CARAML codebase is accessible at \url{https://github.com/FZJ-JSC/CARAML}. The repository features a clean and straightforward structure, consisting of a \texttt{README.md} file and two main directories: \texttt{llm\_training} and \texttt{resnet50}. By focusing on these two representative benchmark cases, the repository is easy to navigate and deploy, enabling users to quickly gather relevant metrics without unnecessary complexity.  

Our approach is to maximize automation in order to facilitate ease of use, reproducibility, and compactness of the benchmarks. To this end, the benchmark codes themselves are not part of the repository. The repository contains the scaffolding code, that automatically downloads codebases, packages, and  execution containers and sets up the required compute environment. To achieve the outlined level of automation, CARAML relies heavily on the JUBE~\cite{breuerJUBE,luehrsJUBE} automation and benchmarking framework. The reported energy measurements are extracted from hardware counters using the self-developed \texttt{jpwr} framework. 

The used Docker container images are provided by the hardware vendors, containing the respective machine learning frameworks, with additional steps to make them usable for CARAML benchmarks. More details can be found in \autoref{Sec:TechnicalDifficulties}.

\subsection{Benchmark Details}
\subsubsection{LLM Training}

For the LLM training benchmark in CARAML, a GPT decoder model is trained from scratch using a subset of the OSCAR data that is preprocessed using GPT-2 tokenizers. The benchmark for NVIDIA GPUs is curated from a specific commit version of Megatron-LM\footnote{\href{https://github.com/NVIDIA/Megatron-LM/tree/f7727433293427bef04858f67b2889fe9b177d88}{https://github.com/NVIDIA/Megatron-LM}} to make it  compatible with all NVIDIA GPU generations. For AMD devices, the BigCode Project fork\footnote{\href{https://github.com/bigcode-project/Megatron-LM/tree/21045b59127cd2d5509f1ca27d81fae7b485bd22}{https://github.com/bigcode-project/Megatron-LM}} is used, which contains adaptations to utilize \texttt{ROCm}  instead of \texttt{CUDA}. In the case of Graphcore, a forked version of a vendor-provided application example\footnote{\href{https://github.com/chelseajohn/examples}{https://github.com/chelseajohn/examples}} is used. All benchmarks employ \jpwr{} to provide a power measurement feature. This necessitates performing a \mintinline{bash}{git patch} to Megatron-LM after cloning the repository. All these steps are automated by JUBE. The patch further contains fixes to streamline the benchmark's automated execution.

The sizing of the specific network architectures, e.g. in terms of number of layers and parallelization configuration, are performed with the aim to fully utilize the hardware system's capabilities. This means that not all accelerators train the exact same model. Due to the different programming paradigm and having only 4 GC200 IPUs available during creation of the suite, only a 117M parameter GPT decoder LLM was trained on the Graphcore device, instead of the 800M parameter GPT decoder model that is used on NVIDIA and AMD hardware. Further JUBE configurations for models containing 13B and 175B parameters are provided in the suite. They can be executed when necessary resources are available, and were tested on NVIDIA GH200 devices. 

Megatron-LM leverages several optimization features, including flash attention, distributed optimizers, activation recomputation, mixed precision, and rotary positional embeddings, in conjunction with various parallelization strategies such as data, tensor, pipeline, and sequence parallelism. For models with 800M parameters, which fit within a single device on both AMD and NVIDIA hardware, only data parallelism is utilized. For the larger model configurations with 13B and 175B parameters,  tensor, pipeline, and sequence parallelism are also enabled. The parallel implementation is done using \emph{PyTorch Distributed}. The benchmark's throughput is measured in terms of \texttt{tokens/second} which is calculated by dividing \texttt{global\_batch\_size $\times$ sequence\_length} with \texttt{elapsed\_time\_per\_iteration}. The benchmark uses all the possible optimization features like flash attention, rotary positional embeddings, distributed optimizers and mixed precision and is terminated based on the value of the \mintinline{bash}{--exit-duration-in-mins} command line argument in Megatron-LM.  

To work around the limited available memory of the Graphcore IPU, we chose a smaller GPT model size (117M), and further employ pipeline parallelism to distribute the model's layers (including the embedding layer) to four devices, using \texttt{Poplar}~\cite{poplar}. This decreases the memory demand per device. Our evaluated system (IPU-POD4 with four IPU, see \autoref{Tab:Systems}) contains four GC200 IPUs, which means that we use a single replica and a single instance (i.e. no data parallelism). Scaling to more nodes can be done by employing more instances using \texttt{PopDist} and \texttt{Horovod}. It is possible to use synthetic data with the benchmark instead of OSCAR data. The benchmark is executed for one epoch and the throughput is measured again in terms of \texttt{tokens/second}, but calculated by dividing \texttt{global\_batch\_size} with \texttt{elapsed\_time\_per\_iteration}, as the \texttt{global\_batch\_size}  is given in number of tokens and not number of samples. 

\subsubsection{ResNet50 Training}

The CARAML ResNet50 benchmark for NVIDIA and AMD is curated from the forked version of official TensorFlow benchmarks\footnote{\href{https://github.com/chelseajohn/tf_cnn_benchmarks}{https://github.com/chelseajohn/tf\_cnn\_benchmarks}}. The main addition in the forked version is the power measurement using \jpwr{} (see section \autoref{Sec:jpwr}). The benchmark uses the \texttt{ResNet50} model, but other models like \texttt{inception3}, \texttt{vgg16}, and \texttt{alexnet} can also be utilized. The benchmark trains a \texttt{ResNet50} model from scratch for 100 iterations and outputs the throughput in \texttt{images/second} computed by dividing the \texttt{global\_batch\_size} by \texttt{elapsed\_time\_ per\_iteration}. Training data can be passed as an argument to the benchmark, or else synthetic data is used. The benchmark is scaled to multiple GPUs using data parallelism implemented with \texttt{Horovod}. The benchmark uses mixed precision and the \texttt{openXLA}~\cite{openxla} compiler for accelerating training.

When targeting Graphcore devices, CARAML uses a forked version of vendor-provided application examples, similar to LLM training, that incorporated power measurement using \jpwr{} (see section \autoref{Sec:jpwr}. The benchmark uses the \texttt{ResNet50} model, but \texttt{ResNet18} and \texttt{ResNet34} models can also be executed with modified configuration files. Similar to the TensorFlow benchmark, used for NVIDIA and AMD devices, a \texttt{ResNet50} model is trained from scratch for one epoch and \texttt{images/second} is used as the throughput metric. Using the \texttt{Poplar} library, provided by Graphcore, the benchmark can be scaled to multiple IPUs using data parallelism when using a single instance; for multiple instances, \texttt{PopDist} and \texttt{Horovod} are used. It is possible to pass training data as an argument or use synthetic data generated either on the host CPU and transferred to the IPU or generated directly on the IPU. Mixed precision training and other custom device optimizations like memory and device mapping, 8-bit transfers, and fused preprocessing are used to make the training efficient.

\subsubsection{Automation with JUBE}
The JUBE~\cite{breuerJUBE,luehrsJUBE} workflow environment facilitates reproducibility and ease of use of the provided benchmarks. Each of the two benchmarks is fully characterized by configuration files, called \emph{JUBE scripts}, where hyperparameters and execution steps are defined. 

A JUBE script can be in \texttt{XML} or \texttt{YAML} format. 
For illustrative reasons, we provide the scripts for LLM training in \texttt{YAML} (\texttt{llm\_training/llm\_benchmark\_nvidia\_amd.yaml} and \texttt{llm\_training/llm\_benchmark\_ipu.yaml}) and the script for training the image classification model in \texttt{XML}  
(\texttt{resnet50/resnet50\_benchmark.xml}).

The execution steps include downloads, compilation, training, and verification. Different systems and steps are executed by supplying the required \emph{tags}. The JUBE runtime interprets the script, resolves dependencies and submits jobs to the Slurm batch system. The job templates are populated from a system-specific configuration file, \texttt{platform.xml}, making the approach system-agnostic. JUBE presents the benchmark results, including a throughput figure-of-merit (\texttt{images/second} and \texttt{tokens/second}) along with energy consumed per device in Watt hour (Wh) during the course of the model training in the benchmark, in compact tabular form after execution.

The JUBE scripts can be utilized to define a set of experiments aimed at exploring the impact of various parameters on performance, such as batch size, optimizers, and learning rate. JUBE simplifies the process of conducting model layout and scaling experiments by automatically generating job scripts with different parameter permutations. Beyond machine learning hyperparameters, this exploration can be extended to system-level configurations, including number of CPU cores or threads, CPU binding strategies and accelerator affinity in terms of NUMA domains.

\subsubsection{Power Measurements with \jpwr{}}
\label{Sec:jpwr}
\jpwr{} is a modular tool for measuring power and energy of different compute devices, currently supporting methods for querying AMD and NVIDIA GPUs, as well as specific methods for getting system power measurements from NVIDIA Grace-Hopper chips and Graphcore IPUs. The code is available at \url{https://github.com/FZJ-JSC/jpwr/} under an AGPL-3.0 license. It can be used either as a command-line tool \texttt{jpwr}, or within Python code as a context manager \mintinline{python}{get_power}. The command line tool wraps other applications, specifying the method to extract power measurements via command line switch, determined by the examined hardware.

The following example shows how \texttt{jpwr} is used to get energy measurements for an application call \mintinline{bash}{stress-ng --gpu 8 -t 5} on an AMD GPU supporting \texttt{ROCm}, writing the results to a CSV file:

\begin{minted}{bash}
jpwr --methods rocm --df-out energy_meas --df-filetype csv stress-ng --gpu 8 -t 5
\end{minted}

The next example shows how the context manager can be invoked for GH200 GPU and system measurements, to save gathered metrics in the object \texttt{measured\_scope}:
\begin{minted}{python}
from jpwr.gpu.pynvml import power
from jpwr.sys.gh import power as gh_power
from jpwr.ctxmgr import get_power
[...]
met_list=[power(),gh_power()]
with get_power(met_list, 100) as measured_scope:
    application_call()
print(measured_scope.df)
\end{minted}

The context manager initiates a power-measurement loop in a separate thread, which periodically queries power consumption using device-specific interfaces, saving data points along with their timestamps. At the end of the operation, these data points are used to calculate the total amount of energy consumed.  The device-specific interfaces, referred to as ``methods'', are implemented as individual modules that can be passed to the context manager.

As backends, vendor-provided libraries and a \texttt{sysfs} interface are employed to extract hardware counters. NVIDIA GPUs use \texttt{pynvml}~\cite{pynvml}, which provides bindings for the NVIDIA Management Library, which is also used by the popular NVIDIA System Management Interface (\texttt{nvidia-smi}). For AMD GPUs, we use the Python module \texttt{rsmiBindings}~\cite{pyrsmi}, which is shipped with the ROCm System Management Interface (\texttt{rocm-smi}). Graphcore IPUs are queried using the Graphcore IPU Info library (\texttt{gcipuinfo}\cite{ipuinfo}), which is also available as a Python module. To also include CPU metrics for GH200 CPU/GPU superchips, the Linux kernel's \texttt{sysfs} interface is used by reading data from device files under the path \texttt{/sys/class/hwmon/}~\cite{ghtuningguide} (called \texttt{gh} in the tool). Multiple backends can be used at the same time, which is useful for GH200, where both \texttt{pynvml} and \texttt{sysfs} methods can be used, or in exotic systems with multiple types of accelerator. The modular structure of these methods ensures they are easily maintained and allows for the seamless addition of further interfaces.

\jpwr{} saves the measured data as Pandas DataFrames internally and this data can be exported. For the command-line tool, specifying \mintinline{bash}{--df-out} and \mintinline{bash}{--df-filetype} arguments sets the output directory and filetype for the DataFrames accordingly. The tool will save all available power and energy data in the specified directory using the specified filetype (HDF5's \texttt{.h5} or \texttt{.csv}). For the context manager, the \texttt{measured\_scope.df} DataFrame contains the power measurement data, and \mintinline{python}{energy_df, additional_data = measured_scope.energy()} returns an energy DataFrame derived from the measurement data in \texttt{energy\_df} and a dictionary of additional DataFrames in \texttt{additional\_data}.

The tool works per-node, i.e. for an MPI or other types of multi-node applications, writing the result files would result in a race condition. To combat this, the tool allows adding a suffix to all result files with the \mintinline{bash}{--df-suffix} option. Furthermore, the suffix string can contain a \mintinline{bash}{

\subsection{Benchmark Execution}\label{sec:caraml_exec}

After cloning the CARAML repository, each benchmark can be executed with just a few JUBE commands, providing the desired benchmark's JUBE script and a tag to define the target architecture. The system tags can be found in the overview of considered systems in \autoref{Tab:Systems}.

For the LLM training benchmark, the required system and model parameters are to be set in \texttt{llm\_training/llm\_benchmark\_nvidia\_amd.yaml} (for NVIDIA and AMD systems) or \texttt{llm\_training/llm\_benchmark\_ipu.yaml} (for Graphcore). 

For the ResNet50 benchmark, the required system and model parameters and the path to the downloaded ImageNet data need to be set in  \texttt{resnet50\_benchmark.xml}. 

More details can be found in Appendix~\ref{appendix:a}.

\section{Results}\label{Sec:Results}
In the following, we report throughput measurements obtained from the hardware systems described in \autoref{Tab:Systems} alongside the corresponding energy consumption data collected during the execution of the CARAML benchmarks. The benchmarks were conducted with careful consideration of CPU binding, MPI threading, and GPU affinity to ensure optimal conditions on the examined machines.

\subsection{LLM Training}

\begin{figure*}[h]
  \centering
\includegraphics[width=\textwidth]{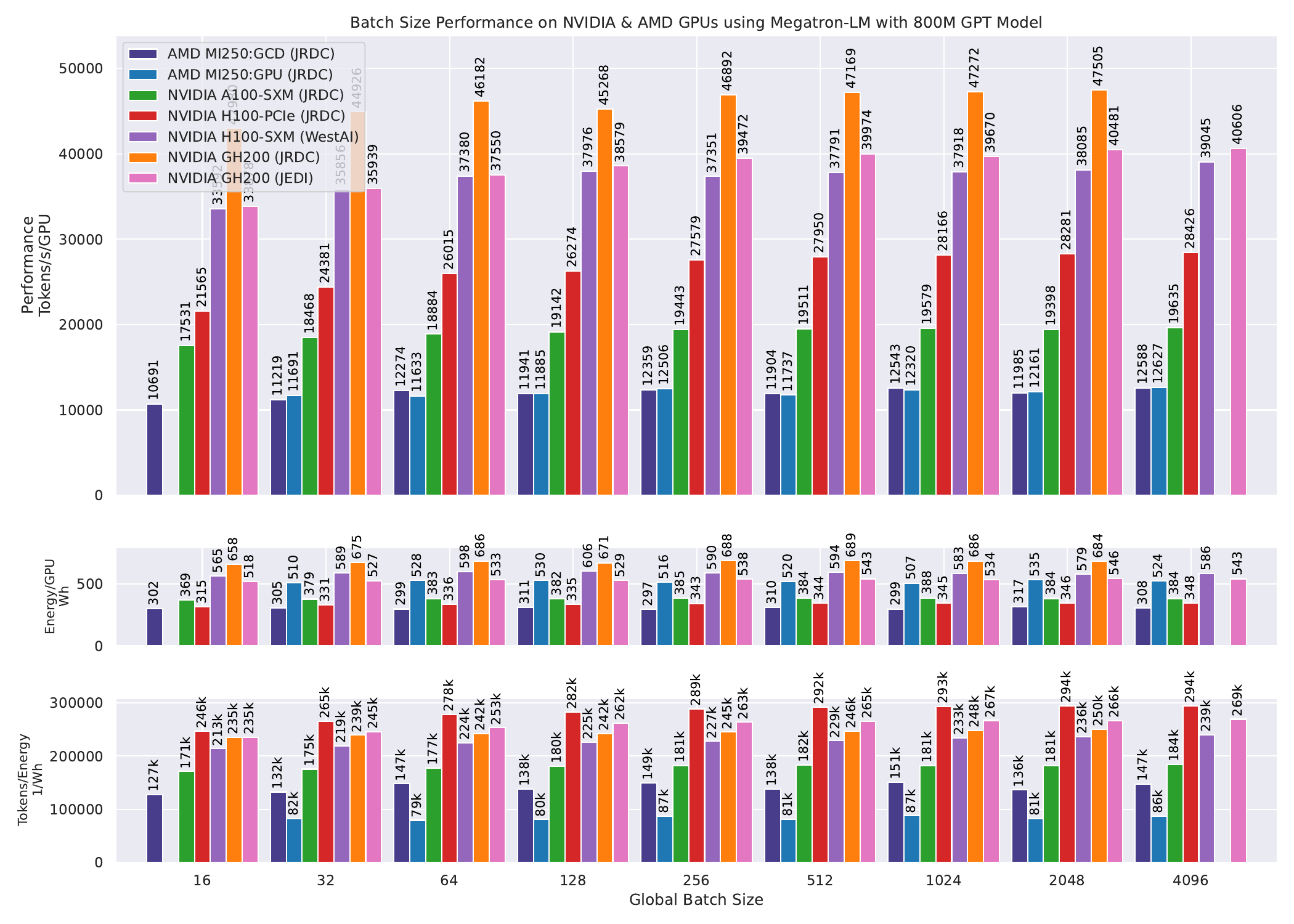}
  \caption{Throughput and energy efficiency for LLM training on NVIDIA and AMD systems using a 800M GPT model.}
  \label{Fig:LLMNVIDIA}
\end{figure*}

\autoref{Fig:LLMNVIDIA} provides throughput results in \texttt{tokens/second} per GPU for NVIDIA systems of various generations and the AMD MI250 GPU, for global batch sizes from 16 to 4096. All experiments train a decoder-only transformer model with 800M parameters using a subset of the OSCAR dataset preprocessed using GPT-2 tokenizers. Since the 800M model fits on a single device, data parallelism can be employed to scale the model across multiple accelerators within a node. The model was trained on an entire node for each system, utilizing data parallelism and micro-batch-size of 4, when multiple accelerators were available. All systems contained 4 GPUs, except the GH200 node in JURECA which has only one (see \autoref{Tab:Systems}). 

For the AMD node, we report two sets of results to draw a complete picture and make a nuanced comparison possible in the context of Multi-Chip Modules (see details in \autoref{Tab:Systems}). The first set of results (\emph{AMD MI250:GCD}) uses 4 GCDs (2 GPUs) with data parallelism of 4, while the second set (\emph{AMD MI250:GPU}) uses all 8 available GCDs (4 GPUs) with data parallelism of 8. When using data parallelism of 8 the global batch size of 16 is not possible since it is not divisible by micro-batch-size times data parallel. In each case, the data is normalized per data parallel (i.e. by 4 and 8, respectively).
Additionally, we present the average total energy consumed per GPU during one hour of model training, along with an energy efficiency metric, calculated as the number of tokens processed per unit of energy consumed.

In general, one can see the performance improvements in more recent GPU hardware generations, with GH200 nodes yielding a throughput of up to  \num{47505} Tokens/s/GPU, $2.45\times$ higher than throughput achieved on A100 GPU nodes. This can be alluded to having more cores and SMs, faster CPU-to-GPU-NVLink connection, TDP, and fast memory. It is evident, that choosing a larger batch size is beneficial for throughput. However, when training a neural network in a production setting, this increased GPU utilization must be balanced against the potential drawback of slower convergence, which could impact the overall training efficiency of the neural network.

Different variants of accelerators were examined, namely the H100 incorporated in the JURECA evaluation platform (referred to here as JRDC) and the H100 in the WestAI cluster, as well as the GH200 in JEDI and the GH200 in JRDC (see Table \ref{Tab:Systems}), and different results can be inferred.

When comparing the two GH200 configurations, we see that a device on a single-accelerator node (GH200 (JRDC)) yields a \qty{20}{\percent} higher performance than a device on a multi-accelerator node (GH200 JEDI)), accompanied by a \qty{20}{\percent} higher energy consumption. Hence, the \texttt{tokens/energy} efficiency per device is similar; even slightly better for the less performant JEDI case.  On JEDI, all devices engage in data-parallel model training, and the additional communication overhead, together with the lower amount of CPU memory available per device, could be the reason for the lower performance per device.  

Another large throughput difference can be observed between the two H100 systems, with the WestAI variant processing  $1.3\times$ as many tokens as the JRDC variant. This could be due to the higher available bandwidth from NVLink connections between GPUs and the SXM GPU form factor, which comes with a higher power envelope (TDP). 

For AMD, using 4 GCDs (2 GPUs) performs slightly better per device than using 8 GCDs (4 GPUs), again representing the overhead of higher parallelization. This overhead leads to a higher energy consumption per device and lower energy efficiency when using 8 GCDs (4 GPUs).

In terms of energy efficiency, the results indicate that the H100-PCIe (JRDC) outperforms all other devices by up to \qty{25}{\percent}, even against the newer technology of GH200 chips, which  provide a throughput twice as high. This is likely related to the limited power budget of the PCIe card, moving its operation mode to a more power-efficient spot. Another factor could be that the other H100 variants, especially the GH200s, are not yet completely saturated in the examined benchmark scenario, as they have higher SM counts. 

\autoref{Tab:LLM_ipu} provides performance and energy efficiency results for the Graphcore machine. Here, the vendor benchmark specifies IPU POD16 as the minimum requirement for GPT-2 PyTorch model training~\cite{ipugpt2readme}. As we only have access to an IPU POD4, a smaller GPT model with 117M parameters is used to benchmark the hardware with energy measurements for global batch sizes from \num{64} to \num{16384}. The larger batch sizes may not be practical for model convergence, but were investigated to understand the limitations of the system. The model layers are split across 4 IPUs and trained for one epoch (global batch size samples) using synthetic data. 
We see in \autoref{Tab:LLM_ipu} that the throughput (\texttt{tokens/second}) increases with the batch sizes, saturating the accelerator, and uses a maximum of \qty{33}{\watt\hour}. The performance is very low compared to GPUs but can partially be explained by the required pipeline parallelism. This form of parallelism introduces a pipeline bubble\cite{narayanan2021efficientlargescalelanguagemodel} and is not as efficient as data parallelism.

\begin{table}
    \addtolength{\tabcolsep}{-0.35em}
    \centering
    \caption{Performance and energy consumption data for training a 117M GPT model for one epoch on IPU GC200 in M2000 POD4. Units for the entries are given in the second row.}
    \begin{tabularx}{\linewidth}{cccc}
    \toprule
    \bf Batch Size &  \bf Tokens/Time & \bf Energy/Epoch/IPU & 	\bf Tokens/Energy \\ 
    & 1/s & Wh & 1/Wh \\
    \midrule
    64    & 64.99 &  15.68 & 4.08 \\
    128   & 97.21 &  18.20 & 7.03 \\
    256   & 129.96 & 18.37 & 13.93 \\
    512   & 155.72 & 18.56 & 27.60 \\
    1024  & 172.94 & 19.07 & 53.71 \\
    2048  & 183.37 & 20.05 & 102.13 \\
    4096  & 188.88 & 21.88 & 187.22 \\
    8192  & 191.86 & 25.47 & 321.34 \\
    16384 & 193.41 & 33.00 & 496.43 \\
    \bottomrule
    \end{tabularx}
    \label{Tab:LLM_ipu}
\end{table}

\subsection{ResNet50 Training}
The ResNet50 training benchmark was performed on all available systems with global batch sizes 16 to 2048.

\begin{figure*}[h]
  \centering
\includegraphics[width=\textwidth]{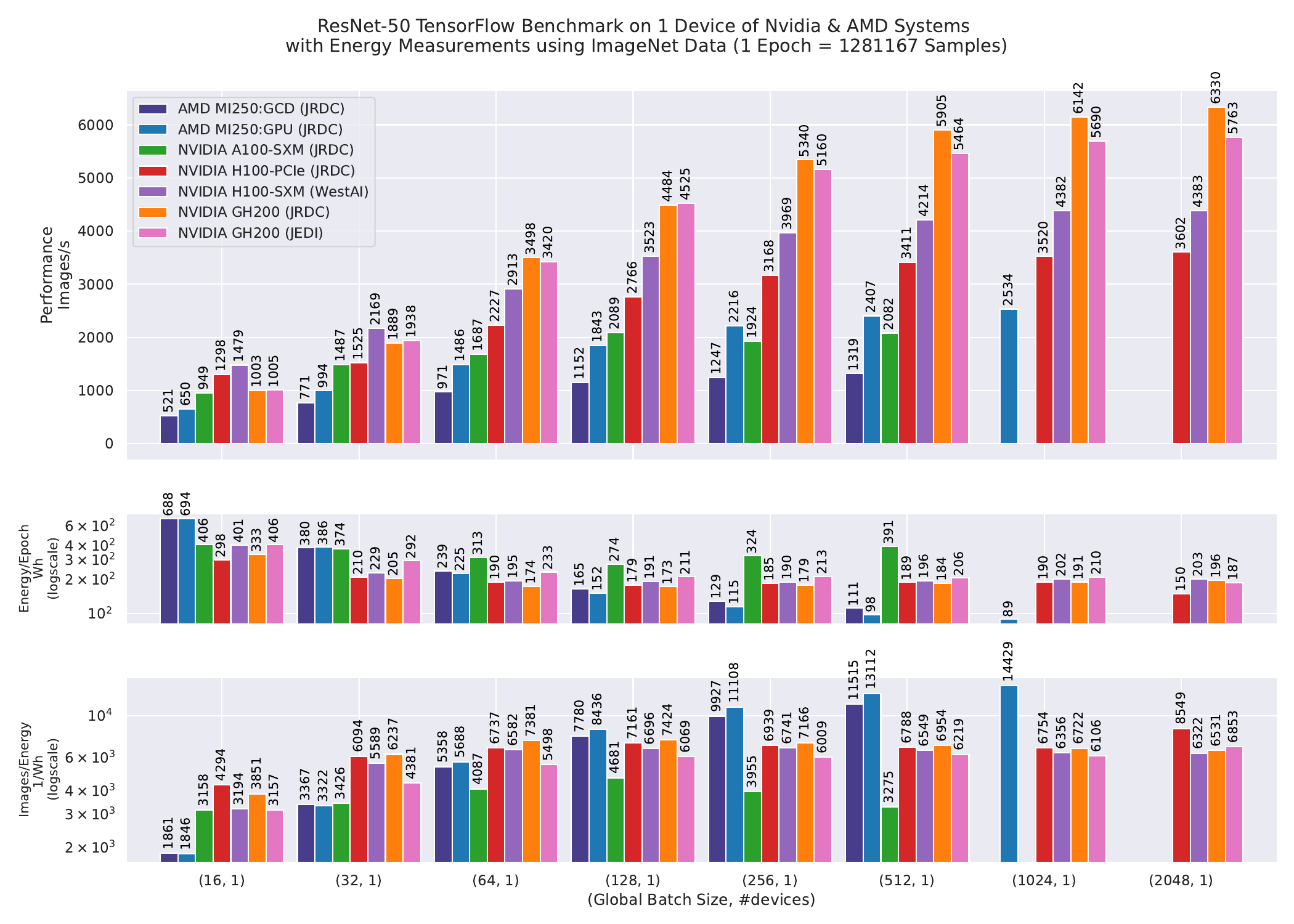}
  \caption{Throughput and energy consumption for \texttt{ResNet50} model training on a single device of NVIDIA and AMD systems.}
  \label{Fig:ResultsResnet}
\end{figure*}

\autoref{Fig:ResultsResnet} reports the throughput of the ResNet50 training process in images per second on a single device on various systems (see \autoref{Tab:Systems}), as well as consumed energy for the whole epoch (processing all images of the input dataset once), and energy efficiency in \unit{images\per\watt\hour}. ImageNet data was used as input, containing \num{1281167} images. 

As expected, the performance increases from older to newer GPU generations. Similar to the LLM training benchmark (see \autoref{Fig:LLMNVIDIA}), we see the WestAI H100 node, with higher-TDP SXM form factor, outperforming the PCIe variant when considering performance per node, with similar energy footprints. Again, GH200 (JRDC) performs better than GH200 (JEDI). This holds true especially for larger batch sizes, which can likely benefit from $4\times$ as much available CPU memory per GPU, allowing for faster data loading. As seen for \texttt{ResNet50} before, the PCIe-variant of the H100 (JRDC), appears to have the best energy efficiency amongst the NVIDIA GPUs, closely followed by the GH200 (JRDC). The reasons seem to stem from a combination of TDP, memory capacity, and bandwidth.

To provide data for more nuanced comparisons in the context of MCMs, as already explained for the LLM results, also for \texttt{ResNet50}, two benchmark runs on the AMD MI250 node were conducted. One run (AMD MI250:GPU) utilized 1 GPU (2 GCDs), requiring data parallelism of 2, and another run (AMD MI250:GCD) utilized only 1 GCD, without parallelism.  Using 2 GCDs naturally leads to a higher throughput, and the device is used more efficiently. This leads to slightly lower amounts of energy needed to process the whole dataset, and a slightly higher energy efficiency. 

The AMD MI250 gives the best efficiency in terms of images per unit of energy for higher batch sizes, while for smaller batches the H100 and GH200 (JRDC) devices are more energy efficient.

\begin{table}
    \addtolength{\tabcolsep}{-0.2em}
    \centering
    \caption{Performance and energy data for training a \texttt{ResNet50} model for one epoch on a single IPU GC200 in M2000 POD4. Units for the entries are given in the second row.}
    \begin{tabularx}{\linewidth}{cccc}
    \toprule
    \bf Batch Size & \bf Images/Time & \bf Energy/Epoch & \bf Images/Energy \\ 
    & 1/s & Wh & 1/Wh \\
    \midrule
    16   & 1827.72 & 32.09 & 39925.87 \\
    32   & 1857.90 & 31.73 & 40382.19 \\
    64   & 1879.29 & 31.75 & 40346.18 \\
    128  & 1888.11 & 31.67 & 40452.50 \\
    256  & 1887.23 & 31.58 & 40563.65 \\
    512  & 1891.74 & 31.49 & 40689.85 \\
    1024 & 1893.07 & 31.50 & 40668.79 \\
    2048 & 1889.87 & 31.53 & 40636.28 \\
    4096 & 1891.58 & 31.51 & 40660.14 \\
    \bottomrule
    \end{tabularx}
    \label{Tab:resnet_ipu}
\end{table}

In the case of Graphcore, the vendor-based TensorFlow \texttt{ResNet50} model training benchmark contains optimizations catering to the IPU execution strategy. When running the benchmark for a single epoch, the IPU first compiles an optimized model graph, which takes close to an hour. It is excluded from the timings presented here. The compiled model graph upon execution is able to complete an epoch with \num{1281167} samples in 10 to 15 minutes. 

\autoref{Tab:resnet_ipu} provides results on throughput, energy consumption, and energy efficiency for the \texttt{ResNet50} benchmark on a Graphcore GC200 IPU. The model performance does not scale on increasing the global batch size. This is likely related to the limitation of not being able to process a micro-batch-size of more than 16 due to limited on-chip RAM (SRAM) and having to execute multiple sequential calls to fetch data from the chip-external memory (DRAM). The energy efficiency compared to classical GPUs looks very promising.

\begin{figure*}
     \centering
     \begin{subfigure}{0.49\textwidth}
         \centering
         \includegraphics[width=\textwidth]{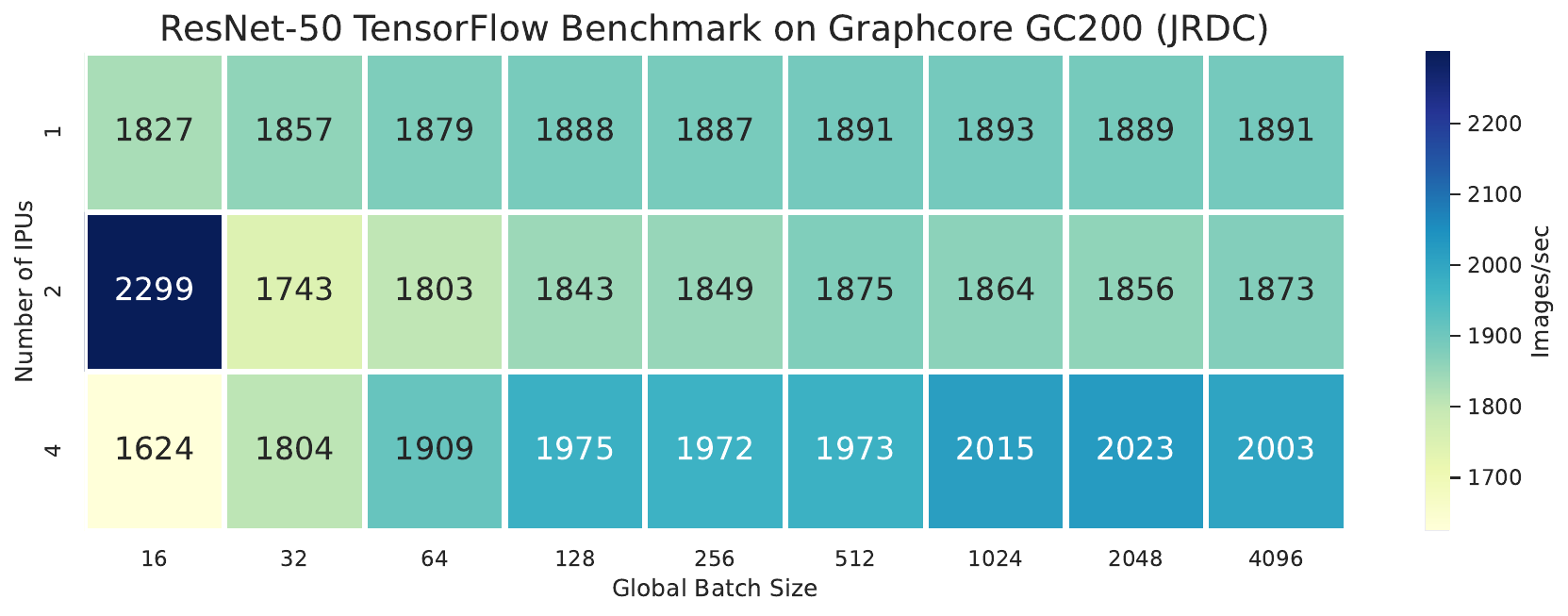}
         \caption{JURECA IPU-M2000}
         \label{Fig:HeatmapGraphcore}
     \end{subfigure}
     \begin{subfigure}{0.49\textwidth}
         \centering
         \includegraphics[width=\textwidth]{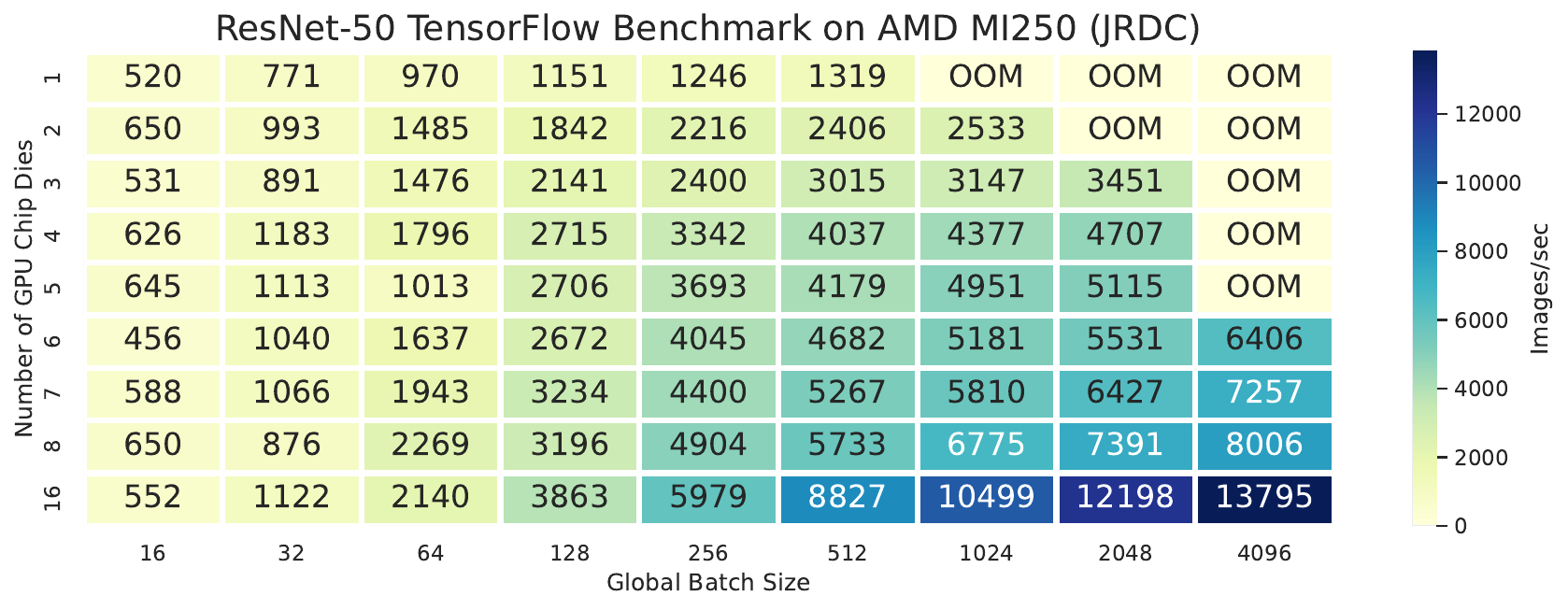}
         \caption{JURECA MI200}
         \label{Fig:HeatmapMI200JURECA}
     \end{subfigure}
     
     \vspace*{0.8\baselineskip}
     
     \begin{subfigure}{0.49\textwidth}
         \centering
         \includegraphics[width=\textwidth]{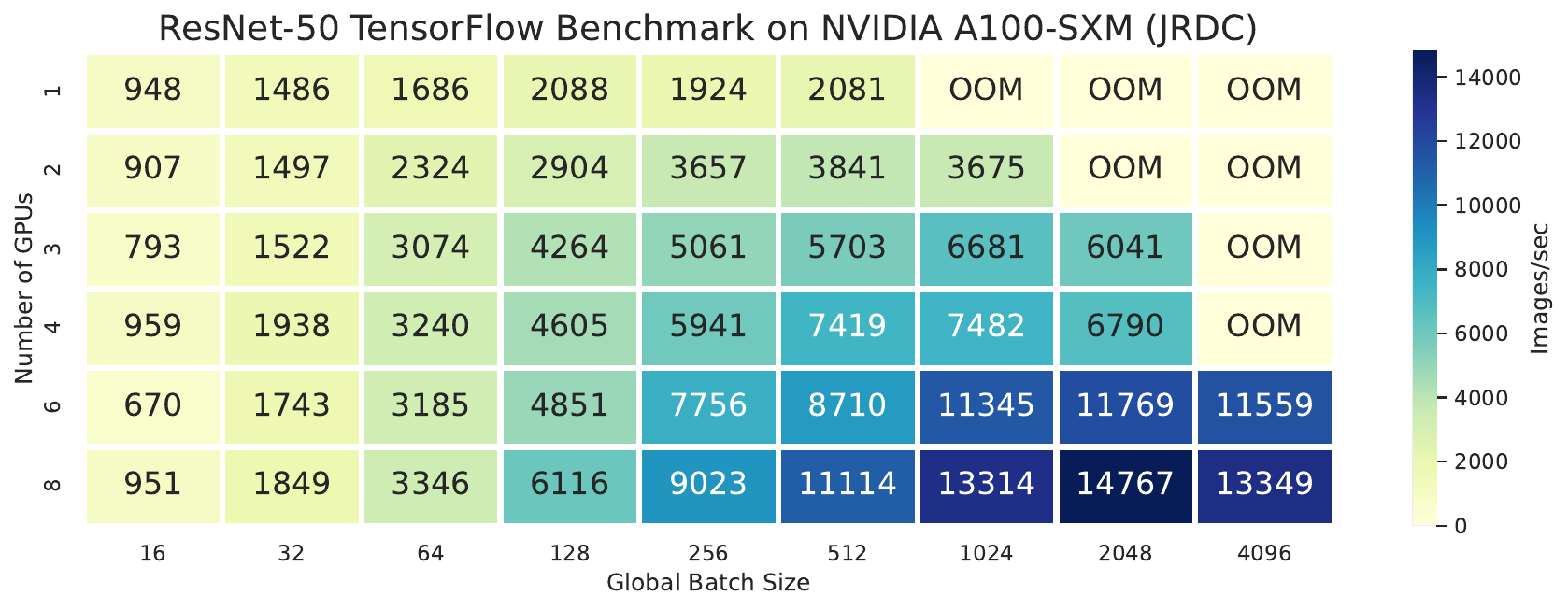}
         \caption{JURECA A100-SXM}
         \label{Fig:HeatmapA100}
     \end{subfigure}
     \begin{subfigure}{0.49\textwidth}
         \centering
         \includegraphics[width=\textwidth]{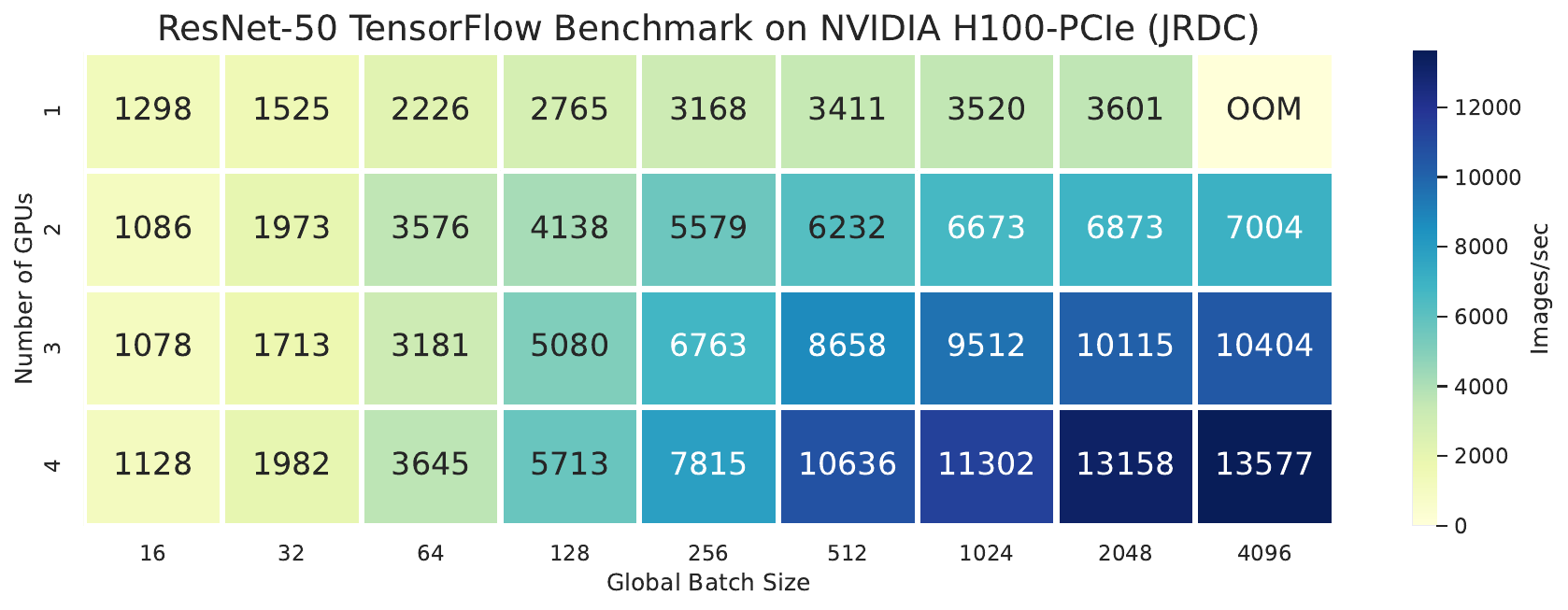}
         \caption{JURECA H100-PCIe}
         \label{Fig:HeatmapH100JURECA}
     \end{subfigure}
     
     \vspace*{0.8\baselineskip}
     
     \begin{subfigure}{0.49\textwidth}
         \centering
         \includegraphics[width=\textwidth]{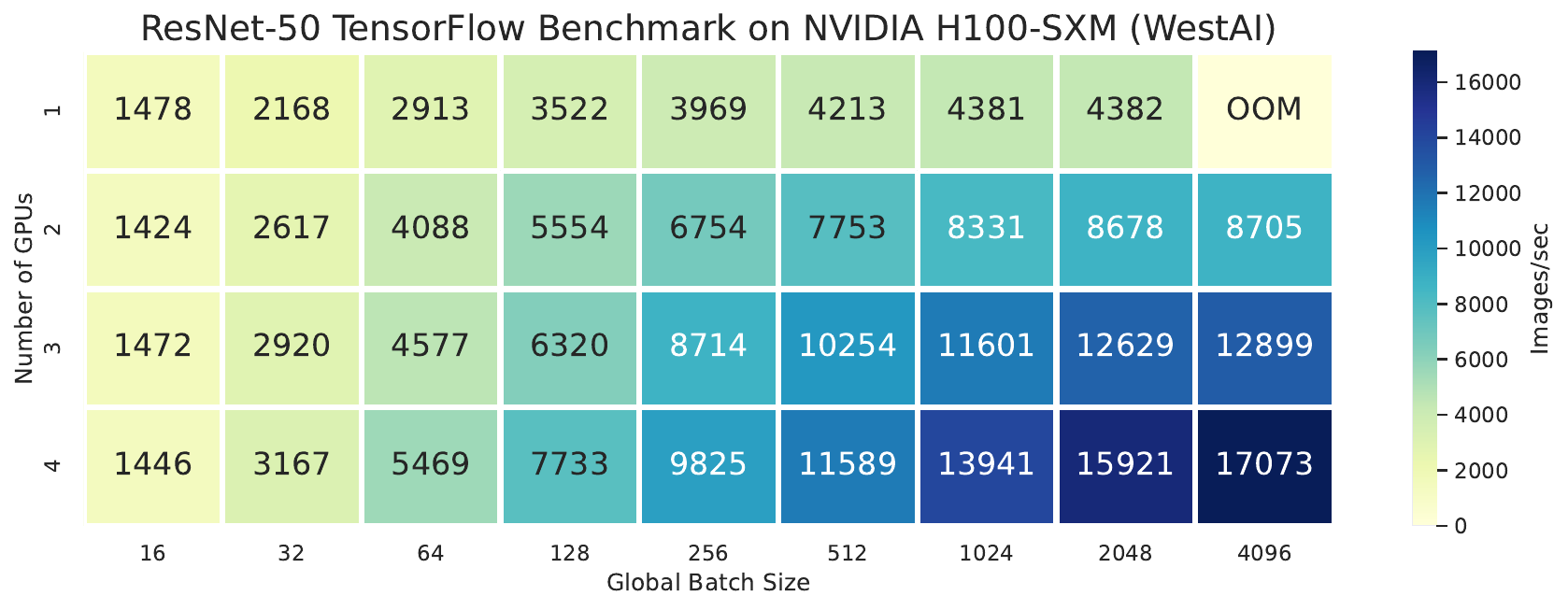}
         \caption{WestAI H100-SXM}
         \label{Fig:HeatmapH100WestAI}
     \end{subfigure}     
     \begin{subfigure}{0.49\textwidth}
         \centering
         \includegraphics[width=\textwidth]{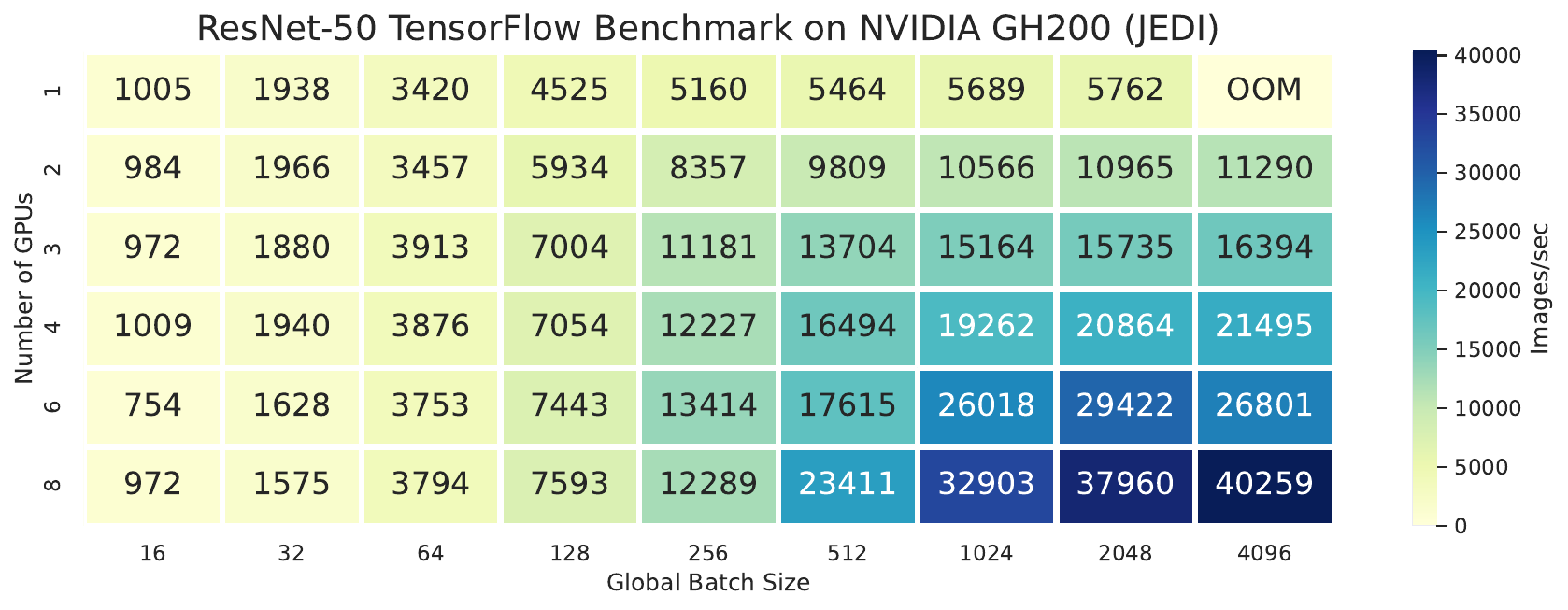}
         \caption{JEDI GH200}
         \label{Fig:HeatmapJEDI}
     \end{subfigure}
     
     \vspace*{0.8\baselineskip}
     
     \begin{subfigure}{0.49\textwidth}
         \centering
         \includegraphics[width=\textwidth]{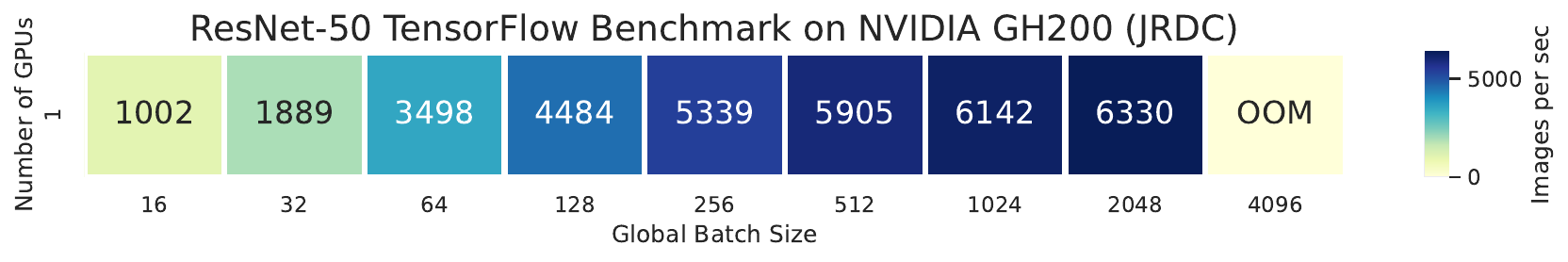}
         \caption{JURECA GH200}
         \label{Fig:HeatmapGH200JURECA}
     \end{subfigure}
     \caption{Throughput for \texttt{ResNet50} training depending on number of GPUs and global batch size on various systems. \emph{OOM} stands for Out of Memory, i.e. the batch size is too large for the memory of the device.}
     \label{fig:heatmaps}
 \end{figure*}

The heatmaps in \autoref{fig:heatmaps} (\ref{Fig:HeatmapGraphcore} to \ref{Fig:HeatmapGH200JURECA}) explore the impact of data-parallelism and global batch size, giving throughput results for all examined architectures. These extensive results for various configurations can be used to estimate the training time required to train a \texttt{ResNet50} model efficiently. 

The heatmaps also contain multi-node results for systems where resources were available. The throughput increases with the global batch size as expected. In the case of Graphcore, the highest throughput was obtained using 2 IPUs for a global batch size of 16, this can be explained due to the batch size fitting into the on-chip RAM, and using fewer IPU links for data transfer.

Judging the results overall, it can be seen that GPUs tend to perform better for larger batch sizes and number of GPUs. In nearly all GPU cases, the best value achieved is for the largest batch size using most GPUs, indicating that the GPUs are not yet saturated, and the limiting factor is the available memory. For Graphcore IPUs performance behavior is relatively flat over a large range of the parameter space, and is best when the batch size fits into the on-chip RAM and multiple calls to DRAM is avoided.

\section{Technical Challenges}
\label{Sec:TechnicalDifficulties}
\subsection{Software}

In CARAML, benchmarks are curated to compare the accelerators with minimal parameter adjustments using open source codebases. For this purpose, NVIDIA and AMD GPUs are tested using the same baseline code. The distinct execution strategy of Graphcore IPUs necessitates a separate codebase, posing a challenge to ensure comparable hyperparameters between IPUs and GPUs. 

The NVIDIA Megatron-LM code base incorporates hardware specific optimizations, with the current version focusing on the Hopper architecture, for example making use of its \emph{Transformer Engine}. To ensure compatibility with the Ampere architecture, CARAML rolls back to a Megatron-LM commit that can be executed on all devices without losing other performance optimizations.

Typically, new optimizations, such as flash attention, are first made available on NVIDIA hardware, with AMD accelerators receiving support later. Currently, the \texttt{flash-attention2} implementation in \texttt{ROCm} is still under development and supports head dimensions only up to 128, whereas the CUDA implementation on NVIDIA GPUs already supports head dimensions up to 256 and \texttt{flash-attention3}.

\subsection{Containers}
Using containers for reproducible workloads promises to simplify the creation of reproducible environments, but setting them up in performance-sensitive HPC environments can be challenging. Issues include locating vendor-supported containers for specific software versions and managing conflicting package dependencies within the container, often with limited permissions. 

Solving these challenges required multiple iterations of container testing, finally leading to the creation of custom containers, inheriting from vendor-provided containers. To manage the installation of additional packages within the container, we utilize \texttt{pip} with the \texttt{--prefix}, \texttt{--no-deps}, and \texttt{--ignore-installed} options, and manually adjust the \texttt{PYTHONPATH}. The container's isolation from the system environment necessitates defining custom  bind paths and the development of wrapper scripts to export environment variables.

Utilizing the shared resources of HPC systems requires usage of job schedulers (Slurm) and message transport libraries (MPI, NCCL). As the employed containers need to bring their own MPI installation, some effort needs to be taken to align the out-of-container distribution setup with the in-container installation. For our setup, the involved PMIx configurations need to be explicitly made compatible by manually setting \mintinline{bash}{PMIX_SECURITY_MODE=native} out-of-container, but within a Slurm-distributed job\footnote{\mintinline{bash}{srun env PMIX_SECURITY_MODE=native apptainer ...}}.

\subsection{System}

In order to ensure a smooth user experience, some non-trivial, HPC-related technical fixes have to be implemented. We document them here to highlight on system-specific issues, which can be helpful when adding more system support for future benchmarks.

As with many HPC systems, the systems at  Jülich Supercomputing Centre feature a high-speed interconnect (InfiniBand) between the nodes. IP connectivity is only available using InfiniBand devices (IP over Infiniband, IPoIB). \texttt{PyTorch} needs to be made aware of the different format of the hostname, which contains an appended \texttt{i} to the \mintinline{bash}{MASTER_ADDR} variable\footnote{Coincidentally, the \texttt{ib0} interface is ordered after the \texttt{en0} interface, such by default the wrong interface is picked. For Jülich Supercomputing systems, the hostnames of IPoIB network are equal to the \texttt{en0}-network hostnames with an appended \texttt{i}.}. Further, a fixed \texttt{torchrun.py} script is required for execution on such a system. While not yet being fixed in the main \texttt{PyTorch} codebase\footnote{The issue is open upstream since 2022: \url{https://github.com/pytorch/pytorch/issues/73656}; meanwhile a patched version is provided on PyPI for convenience, \url{https://pypi.org/project/torchrun-jsc/}.} a patched version is available  in the benchmark repository (\texttt{llm\_training/aux/fixed\_torch\_run.py}). 

With AI-distributed training often depending on external software like \texttt{PyTorch} and \texttt{Horovod}~\cite{sergeev2018horovodfasteasydistributed} to spawn multiple processes, errors due to conflicting MPI ranks with Slurm are encountered at times and have to be fixed.

Further, during the benchmarking process, the critical impact of correct CPU binding, optimal number of threads, and GPU affinity on performance for each system was carefully studied. It was found that a GPU-centric approach to affinity is useful, creating one Slurm task per GPU and distributing them to CPU cores with affinity to respective GPUs. At the same time, it is important to create CPU masks that are open enough for NCCL to place its helper thread. 

JEDI, as one example, features four Grace-Hopper superchips, so that the Slurm options \mintinline{bash}{--ntasks=4 --cpus-per-task=72 --gpus-per-task=1} give the proper affinity. JURECA A100 nodes, as another example, feature EPYC processors in which not all CPU chiplets have GPU affinity. Due to this, explicitly targeting the \emph{proper} NUMA domains with \mintinline{bash}{--cpu-bind} is a complex, but useful approach.

\section{Conclusions}\label{Sec:Conclusions}
As AI continues to experience rapid growth and the market is seeing a growing influx of AI accelerators, evaluating accelerator performance using real world applications is crucial. In this paper, we introduced CARAML, a benchmark suite designed to assess AI workloads on accelerators with energy measurements. CARAML uses the JUBE framework to create compact, automated benchmarks for both LLM and Computer Vision training. The benchmarks incorporate the modular \jpwr{} tool to measure energy consumption. CARAML is further capable to perform ablation studies to identify  hardware and model configurations for optimal performance. The details of the framework and results obtained using CARAML on seven different accelerators systems from NVIDIA, AMD and Graphcore that differ either in generation or configuration were discussed in detail.

The results confirm that the latest accelerator generations yield a better performance, but the energy efficiency is influenced by more factors in the hardware and network configuration. The GH200 generally gives the best performance, related to the CPU-to-GPU-NVLink connection, TDP, and fast memory. The PCIe-flavor of the H100 usually gives the best energy-efficiency, a result of operation at an efficient power operating point.

While the surveyed Graphcore accelerator system could not yield a competitive performance to classical GPUs, the results on energy efficiency are very promising, outperforming GPUs in this regard for some benchmarks. This relies on code that is optimized for the execution on an IPU's data-flow architecture, which can yield performance improvements.

Several technical challenges were encountered while automating the CARAML benchmark setup. Solutions required a deep understanding of networking specifics, AI framework backends, and how containers interact with their environment. 

As future work, we plan to further develop CARAML by incorporating continuous benchmarking capabilities and enhancing its usability. We also aim to expand the suite by including additional AI training and inference benchmarks.

\section*{Acknowledgment}
Part of this work was funded by the German Federal Ministry for Economic Affairs and Climate Action (BMWK) through the project OpenGPT-X (project no. 68GX21007D). Additional support was provided by the EuroHPC Joint Undertaking under Grant Agreement 955513, co-funded by the German Federal Ministry of Education and Research (BMBF) under funding reference 16HPC029 through the MAELSTROM project.

Work presented here made extensive use of JURECA-DC, the JURECA-DC Evaluation Platform, the WestAI infrastructure, and the JUPITER enablement platform JEDI, which we greatly acknowledge. 

We would like to express our gratitude to Jan Ebert and Jan Robert Finkbeiner for their valuable insights and discussions on configuring neural network architectures.

\section*{Reproducibility}
The source codes of CARAML and \jpwr{} are available at \url{https://github.com/FZJ-JSC/CARAML} and \url{https://github.com/FZJ-JSC/jpwr}. The results shown in the paper can be reproduced using CARAML by following the instruction entailed in \autoref{sec:caraml_exec} and the corresponding readme files.

\bibliographystyle{IEEEtran}
\bibliography{references}

\appendix
\section{Appendix A}\label{appendix:a}

To execute the CARAML benchmark, clone the CARAML repository and use the corresponding JUBE script and a tag to identify the target architecture. The available system tags are listed in the overview of systems in \autoref{Tab:Systems}

\paragraph{LLM Training}
\begin{itemize}
    \item Set the required system and model parameters in \texttt{llm\_training/llm\_benchmark\_nvidia\_amd.yaml}(for NVIDIA and AMD systems) or \texttt{llm\_training/llm\_benchmark\_ipu.yaml}(for Graphcore)
    \item To pull the required container and build packages, use container tag as:
        \begin{itemize}
            \item NVIDIA A100 and H100 GPUs
            \vspace{3pt}
            \begin{minted}{bash}
            jube run llm_training/llm_benchmark_nvidia_amd.yaml --tag container H100
            \end{minted}
            \vspace{3pt}
            \item NVIDIA GH200 and JEDI GPUs
                \vspace{3pt}
                \begin{minted}{bash}
                jube run llm_training/llm_benchmark_nvidia_amd.yaml --tag container GH200
                \end{minted}
                \vspace{3pt}
            \item AMD MI250
                \vspace{3pt}
                \begin{minted}{bash}
                jube run llm_training/llm_benchmark_nvidia_amd.yaml --tag container MI250
                \end{minted}
                \vspace{3pt}
            \item Graphcore GC200
                \vspace{3pt}
                \begin{minted}{bash}
                jube run llm_training/llm_benchmark_ipu.yaml --tag container 
                \end{minted}
                \vspace{3pt}
        \end{itemize}
    \item To run the benchmark with defined configurations for \texttt{800M} GPT model with tokenized OSCAR data provided with the repository do:
        \vspace{3pt}
        \begin{minted}{bash}
            jube run llm_training/llm_benchmark_nvidia_amd.yaml --tag A100 800M 
        \end{minted}
        \vspace{3pt}

        A100 can be replaced with H100, WAIH100, GH200, JEDI and MI250 for the respective systems and \texttt{800M} can be replaced with \texttt{13B} and \texttt{175B} for systems with available  node resources. 
    \item To run the benchmark with defined configurations for \texttt{117M} GPT model on Graphcore with synthetic data do:
        \vspace{3pt}
        \begin{minted}{bash}
            jube run llm_training/llm_benchmark_ipu.yaml --tag 117M synthetic
        \end{minted}
        \vspace{3pt}
        If tag synthetic is not given, the benchmark will use the tokenized OSCAR data.

    \item To combine the energy data into a single CSV file and post-process results do:
        \vspace{3pt}
        \begin{minted}{bash}
            jube continue llm_training/llm_benchmark_nvidia_amd_run -i last
        \end{minted}
        Or
        \vspace{3pt}
        \begin{minted}{bash}
            jube continue llm_training/llm_benchmark_ipu_run -i last
        \end{minted}
        \vspace{3pt}

    \item To get the final result in tabular form do:
        \vspace{3pt}
        \begin{minted}{bash}
            jube result llm_training/llm_benchmark_nvidia_amd_run -i last
        \end{minted}
        
        Or
        \vspace{3pt}
        \begin{minted}{bash}
            jube result llm_training/llm_benchmark_ipu_run -i last
        \end{minted}
\end{itemize}
\vspace{5pt}

\paragraph{ResNet50 Training}

\begin{itemize}
    \item Set the required system and model parameters and the path to downloaded ImageNet data in \texttt{resnet50\_benchmark.xml}
    \item To pull the required container, use container tag as:
        \begin{itemize}
            \item NVIDIA A100 and H100 GPUs
                \vspace{3pt}
                \begin{minted}{bash}
                    jube run  resnet50/resnet50_benchmark.xml --tag container H100
                \end{minted}
                \vspace{3pt}
            \item NVIDIA GH200 and JEDI GPUs
                \vspace{3pt}
                \begin{minted}{bash}
                    jube run resnet50/resnet50_benchmark.xml --tag container GH200
                \end{minted}
                \vspace{3pt}
            \item AMD MI250
                \vspace{3pt}
                \begin{minted}{bash}
                    jube run resnet50/resnet50_benchmark.xml --tag container MI250
                \end{minted}
                \vspace{3pt}
            \item Graphcore GC200
                \vspace{3pt}
                \begin{minted}{bash}
                    jube run resnet50/resnet50_benchmark.xml --tag container GC200
                \end{minted}
                \vspace{3pt}
        \end{itemize}
    \item To run the benchmark with defined configurations do:
        \vspace{3pt}
        \begin{minted}{bash}
            jube run resnet50/resnet50_benchmark.xml --tag A100
        \end{minted}
        \vspace{3pt}
        Or with synthetic data
        \vspace{3pt}
        \begin{minted}{bash}
            jube run resnet50/resnet50_benchmark.xml --tag A100 synthetic
        \end{minted}
        \vspace{3pt}
        A100 can be replaced with H100, WAIH100, GH200, JEDI, MI250 and GC200 for the respective systems.
    \item To combine the energy data into a single CSV file and post-process the results do:
        \vspace{3pt}
        \begin{minted}{bash}
            jube continue resnet50/resnet50_benchmark_run -i last
        \end{minted}
        \vspace{3pt}
    \item To get the final result in tabular form do:
        \vspace{3pt}
        \begin{minted}{bash}
            jube result resnet50/resnet50_benchmark_run -i last
        \end{minted}
        \vspace{3pt}
\end{itemize}

\end{document}